\documentclass[preprint,showpacs,preprintnumbers,amsmath,amssymb,nofootinbib]{revtex4}
\usepackage{graphicx,epsfig,dcolumn,bm,epic,eepic,float}
\usepackage{amsmath}
\usepackage{latexsym}
\usepackage{color}
\usepackage{cancel}
\usepackage{makeidx,shortvrb,latexsym}
\begin{document}
\unitlength 1 cm
\newcommand{\nn}{\nonumber}
\newcommand{\vk}{\vec k}
\newcommand{\vp}{\vec p}
\newcommand{\vq}{\vec q}
\newcommand{\vkp}{\vec {k'}}
\newcommand{\vpp}{\vec {p'}}
\newcommand{\vqp}{\vec {q'}}
\newcommand{\bk}{{\bf k}}
\newcommand{\bp}{{\bf p}}
\newcommand{\bq}{{\bf q}}
\newcommand{\br}{{\bf r}}
\newcommand{\bR}{{\bf R}}
\newcommand{\up}{\uparrow}
\newcommand{\down}{\downarrow}
\newcommand{\fns}{\footnotesize}
\newcommand{\ns}{\normalsize}
\newcommand{\cdag}{c^{\dagger}}
\title {Folding model analysis of  $ ^{12}C- ^{12}C $ and $ ^{16}O- ^{16}O $ elastic
scattering using the density-dependent \textit{LOCV} averaged
effective interaction}
\author{$M. \; Rahmat$ }
\author{$M. \; Modarres$}
\altaffiliation{Corresponding author, Email: mmodares@ut.ac.ir,
Tel:+98-21-61118645, Fax:+98-21-88004781.} \affiliation{Department
of Physics, University of $Tehran$, 1439955961, $Tehran$, Iran.}
\begin{abstract}
The averaged effective two-body interaction (\textit{AEI}) which can
be generated through the lowest order constrained variational
(\textit{LOCV}) method for symmetric nuclear matter (\textit{SNM})
with the input \textit{Reid}68 nucleon-nucleon potential, is used as
the effective nucleon-nucleon potential in the folding model to
describe the heavy-ion (\textit{HI}) elastic scattering cross
sections. The elastic scattering cross sections of $^{12}C-^{12}C$
and $^{16}O-^{16}O$ systems are calculated in the above frameworks.
The results are compared with the corresponding calculations coming
from the fitting procedures with the input finite range
\textit{DDM3Y1-Reid} potential and the available experimental data
at different incident energies. It is shown that a reasonable
description of the elastic $^{12}C-^{12}C$ and $^{16}O-^{16}O$
scattering data at the low and the medium energies can be obtained
by using the above \textit{ LOCV AEI}, without any need to define a
parameterize density dependent function in the effective
nucleon-nucleon potential, which is formally considered in the
typical \textit{DDM3Y1-Reid} interactions.
\end{abstract}
 \maketitle
\section{Introduction}
In recent years, there has been a growing interest in the heavy-ion
(HI) scattering. These  collision processes were investigated widely
both experimentally and theoretically. One of the goals of studying
the HI reactions is to determine the form of the most suitable
effective nucleon-nucleon potential,   to explain the experimental
elastic scattering cross section data \cite{4,4p}. For many years,
the use of empirical parametrization of nuclear potential was very
common in the HI studies, but it is desirable to relate the
nucleus-nucleus ($ \mathcal{NN} $) interactions to the
nucleon-nucleon (\textit{NN}) nuclear potential \cite{2}. Many
attempts in this direction have been made, and recently, the
double-folding (\textit{DF}) model was extensively used by many
groups in describing the HI scattering, since it gives a simple
possibility of numerical handling in two nucleus scattering
calculations \cite{3}.

In the folding model, the potential is usually generated by folding
an effective \textit{NN} interaction over the ground-state density
distribution of the two nuclei \cite{4,4p}. In general, we need a
well-defined effective \textit{NN} interaction which reproduces the
basic nuclear matter properties (like the saturation energy and
density), and, on the other hand, it can be used as a basic input in
the description of HI scattering qualitatively with respect to the
experimental data   \cite{5}. The \textit{M3Y} interaction \cite{6}
and its density dependent versions \cite{7,8,9,10,11,12,13}, are
usually used into the folding model. Recently the G-matrix and
extended $Hartree-Fock$   approaches \cite{131,132,133,134,135,136}
with  and without the inclusion of the three body force (TBF) and
the rearrangement term (RT), were applied for calculating  the
nucleon-nucleus and the nucleus-nucleus scattering cross-section
calculations (but mainly at 70 $MeV$), as well as   obtaining the
nuclear matter saturation properties ($EOS$) \cite{131}. The RT
comes out in case of calculating the single particle energy and the
corresponding potential. But in the present work, we intend to apply
the lowest order constrained variational averaged effective
interaction \textit{LOCV AEI}, which was generated by using the
input \textit{Reid}68 potential in our previous work \cite{14}, as
the effective \textit{NN} interaction, into the folding model to
test the validity of our interaction in describing the HI elastic
scattering. In this paper, we limit ourselves to the elastic
scattering of spherical projectile and spherical target nuclei, so
we consider the $^{12}C- ^{12}C$ and $ ^{16}O- ^{16}O $ elastic
scattering.

A brief discussion about the $LOCV$ method is given in the appendix
A. Contrary to G-matrix approach, in the $LOCV$ formalism (which is
based on the cluster expansion \cite{clark}), the wave functions,
e.g. the correlation functions, are calculated through the
Euler-Lagrange differential  equations, whereas the application of G
operator on the plane wave generate the interacting wave functions.
Another advantage of the cluster expansion is its expansion in the
powers of correlation functions (in the G-matrix language the wound
parameter) and the first power of the \textit{NN} potential. So it
converges faster than the G-matrix approaches which is an expansion
in the powers of the potential. On the other hand since we directly
calculate the \textit{LOCV AEI},   there is no need to calculate the
RT in our approach. In  the table 1, the results of the $LOCV$
saturation properties of symmetrical nuclear matter ($SNM$)
calculation for the $Reid68$ and $\Delta-Reid68$ potentials, (in
comparison to the empirical one), are presented. The $LOCV$ method
is self-consistently predict the $EOS$ of $SNM$ (for the detail see
the appendix and the table A.1). The one-body, ($E_1$ (is simply the
Fermi energy)), the two-body cluster, ($E_2$), and  the three-body
cluster ,($E_3$), terms as well as  the convergence parameters are
discussed in the appendix.

In some of our $LOCV$ calculations, we have taken into account the
effects of $TBF$ such as the $\Delta$ box diagram (see the appendix
A). But in the present work since we intend to compare our results
with those coming from the \textit{M3Y} interaction \cite{6} which
is based on the \textit{Reid}68 potential, so our results will be
limited to this interaction. However we hope in our future works,
the other interactions as well as the effects of the $TBF$ on the
nucleus-nucleus differential cross sections are evaluated. In the
table A.1 it is clearly demonstrated that the $LOCV$ method predicts
the $SNM$ saturation properties close to other methods, even with or
without $TBF$ \cite{464}. We should point out here that there is no
extra parameters and conditions on the $LOCV$ method to predict the
saturation properties of $SNM$.

In our recent paper \cite{14} , we   derived the averaged effective
two-body interactions (\textit{AEI}) through the lowest order
constrained variational (\textit{LOCV}) calculations for the $SNM$
with the \textit{Reid}68  \cite{15}, the $\triangle$-\textit{Reid}68
\cite{16} (which takes into the account the effect of three-body
force ($TBF$)) and the A$\upsilon_{18}$ \cite{17} interactions as
the input phenomenological nucleon-nucleon potentials, and
reformulated them in the radial and density-dependent parts as well
as its direct and exchange components . Note that the radial parts
are fixed and density dependent functions only depend on density
which becomes a constant at fix density, i.e. similar to the $M3Y$
calculations. Here as we stated above, we only use the \textit{LOCV
AEI} with the input \textit{Reid}68 potential into the folding model
and compare our results with those coming from the
\textit{DDM3Y1-Reid} which uses a finite range potential as the
direct and exchange components i.e. $M3Y$ interactions \cite{3}. The
\textit{LOCV} effective two-body interactions were tested by
calculating the properties of the light and the heavy closed shell
nuclei \cite{18,19,20}, and recently it was used to calculate the
in-medium \textit{nn} cross section, the transport properties of
  neutron matter \cite{21,22} and the normal liquid Helium-3
\cite{23}. In these works, it was shown that the \textit{LOCV}
\textit{AEI} gave the reasonable  results in comparison to the
corresponding available data.

So, this article is organized as follows: In the section 2, we
briefly review the theoretical formalism of the double folding
model. The density distributions and the different kinds of the
effective interactions used into the folding model as well as the
computational procedure are also discussed in this section. Finally,
while the results of the calculations and discussions are given in
the section 3,   the section  4 is devoted to the summary and
conclusions.
\section{THE THEORETICAL FORMALISM }
\subsection{The double folding model}
Satchler and Love \cite{24}   presented the basic idea  of the
folding model in detail and  in the reference \cite{3}, an improved
version of folding model was introduced to calculate the exchange
part of the HI potential. We give here only a brief description of
this model and refer the reader  to the references
\cite{4,4p,24,25,26,27} for details. In the first order of
Feshbach's theory for the optical potential, the microscopic
nucleus-nucleus potential can be evaluated as an antisymmetrized
$Hartree Fock$ type potential for the $dinuclear$ system
\cite{4,4p,3}:
\begin{equation}
U=U_{D}+U_{EX}=\sum_{i\in A_{1} , j\in A_{2}} [\langle
ij|\upsilon_{D}|ij \rangle - \langle ij|\upsilon_{EX}|ji \rangle],
\end{equation}
where $ |i \rangle $ and $ |j \rangle $ refer to the single-particle
wave functions of nucleons in the two colliding nuclei $ A_{1} $ and
$ A_{2} $, respectively; $ \upsilon_{D} $ and $ \upsilon_{EX} $ are
the direct and the exchange parts of the effective \textit{NN}
interaction. After doing some algebra, one can explicitly write the
energy-dependent direct and exchange potentials as,
\begin{equation}
U_{D}(E,\textbf{R})=\int d\textbf{r}_{p}
d\textbf{r}_{t}\rho_{p}(\textbf{r}_{p})\rho_{t}(\textbf{r}_{t})\upsilon_{D}(\rho,E,s)\:,\;\;\textbf{s}=\textbf{r}_{p}-\textbf{r}_{t}+\textbf{R},
\end{equation}
\begin{equation}
U_{EX}(E,\textbf{R})=\int d\textbf{r}_{p}
d\textbf{r}_{t}\rho_{p}(\textbf{r}_{p};\textbf{r}_{p}+\textbf{s})\rho_{t}(\textbf{r}_{t};\textbf{r}_{t}-\textbf{s})\upsilon_{EX}(\rho,E,s)e^{(i\textbf{k}_{rel}.\textbf{s}/A_{red})}.
\end{equation}
Note that, in general the one-body density is written as $
\rho(\textbf{r},\textbf{r}^{\prime}) $. In the case of direct term,
it becomes $ \rho(\textbf{r}_{p}) $ or $ \rho(\textbf{r}_{t}) $,
i.e. the diagonal terms, where $ \textbf{r}_{p} $ and $
\textbf{r}_{t} $ are the positions of the two nucleons in the nuclei
p (projectile) and t (target), respectively, $.
\textbf{s}=\textbf{r}_{p}-\textbf{r}_{t}+\textbf{R} $ corresponds to
the distance between the two specified interacting points of the
projectile and the target, and $ \textbf{R} $ is a vector from the
center of the t nucleus to that of p nucleus. But in case of the
exchange terms, we have $ \rho(\textbf{r},\textbf{r}^{\prime}) $ for
each nucleus, i.e. nondiagonal terms, with ($
\textbf{r}=\textbf{r}_{p} ,
\textbf{r}^{\prime}=\textbf{r}_{p}+\textbf{s} $) or ($
\textbf{r}=\textbf{r}_{t} ,
\textbf{r}^{\prime}=\textbf{r}_{t}-\textbf{s} $). So for the
exchange term the densities are the functions of two different
coordinates \cite{3}. In the above equations, the wave number $
\textbf{k}_{rel} $ associated with the relative motion of colliding
nuclei, which is given by:
\begin{equation}
 k_{rel}^{2}(\textbf{R})=2m_{n}A_{red}[E_{c.m.}-U(E,\textbf{R})-V_{C}(\textbf{R})]/\hbar^{2},
\end{equation}
where  $ A_{red}=A_{p}A_{t}/(A_{p}+A_{t}) $, $ m_{n} $, $ E_{c.m.} $
and $ E $ are  the reduced mass number, the bare nucleon mass, the
center-of-mass (c.m.) energy and   the incident laboratory energy
per nucleon, respectively. Here $
U(E,\textbf{R})=U_{D}(E,\textbf{R})+U_{EX}(E,\textbf{R}) $ and $
V_{C}(\textbf{R}) $ are the total nuclear and the Coulomb
potentials, respectively. It can be seen from  the equation (3) that
the energy-dependent HI potential is nonlocal through its exchange
term. For simplicity of the numeric calculations, a realistic local
expression for the density matrix is usually  used \cite{28}:
\begin{equation}
\rho(\textbf{R},\textbf{R}+\textbf{s})\simeq
\rho(\textbf{R}+\dfrac{\textbf{s}}{2})\hat{j}_{1}(k_{F}(\textbf{R}+\dfrac{\textbf{s}}{2})s),
\end{equation}
where $ \hat{j}_{1}(x)=3(\sin x -x\cos x) /x^{3} $. The explicit
form of $ k_{F}(\textbf{R}) $ is given in the reference \cite{3}. In
order to specify the overlap density during the HI collision, we
have applied the procedure used in the reference \cite{3} that is
called frozen density approximation ($FDA$). In this approach, the
overlap density, $ \rho $, is taken to be the sum of the densities
of the target and the projectile densities at the midpoint of the
inter-nucleon separation, i.e.,
\begin{equation}
\rho=\rho_{p}(\textbf{r}_{p}+\dfrac{\textbf{s}}{2})+\rho_{t}(\textbf{r}_{t}-\dfrac{\textbf{s}}{2}).
\end{equation}
This procedure simply corresponds to the local density approximation
assumed in the different nuclear matter studies \cite{3,18,19,20}.

After performing  some transformations one can obtain the exchange
potential in the following local form:
\begin{eqnarray}
U_{EX}(E,\textbf{R})=4\pi \int_{0}^{\infty }\upsilon_{EX}(s,E)s^{2} ds \hat{j}_{0}(k(\textbf{R})s/M) \nonumber\\
\times\int f_{1}(\textbf{r},s) f_{2}(\textbf{r}-\textbf{R},s)
F[\rho_{p}(\textbf{r})+\rho_{t}(\textbf{r}-\textbf{R})]d\textbf{r},
\end{eqnarray}
where ($F(\rho)$  will be defined later on, i.e. see the equations
(19) to (25) in the subsection II-B),
\begin{equation}
f_{1(2)}(\textbf{r},s)=\rho_{p(t)}(r)\hat{j}_{1}(k_{F1(2)}(r)s)
\:\:\:,\;\;\: \hat{j}_{0}(x)=\dfrac{\sin x}{x}.
\end{equation}
Applying the folding formulas in the momentum space \cite{28}, one
can write the exchange potential as:
\begin{equation}
U_{EX}(E,\textbf{R})=4\pi \int_{0}^{\infty
}G(\textbf{R},s)\hat{j}_{0}(k(\textbf{R})s/M)\upsilon_{EX}(s,E)s^{2}
ds.
\end{equation}
The explicit form of $ G(\textbf{R},s) $ function can be found in
the reference \cite{3}.

As it can be seen from the equation (4), the wave number of relative
motion, $  k_{rel}(R) $, depends on the total HI potential, so, we
encounter with a self-consistency problem in obtaining the exchange
part of HI potential at each radial point. In general, this problem
can be  overcome by applying an iterative procedure, as it was
performed for the first time by Chaudhuri et al. \cite{29}. However,
in the references \cite{26,27} a closed expression was used to
obtain the exchange potential by using the multiplication theorem of
the Bessel function $ \hat{j}_{0}(k(\textbf{R})s/M) $. In this
paper, we use the iterative method to ensure the self-consistency at
all the radial point, in which, we chose $ U_{D}(E,R) $ as the
starting potential to enter in the $ \hat{j}_{0}(k(\textbf{R})s/M) $
term in the exchange integral, the equation (9).

Since the effective \textit{NN} interactions applied into the
folding model are real, the calculated HI potentials are also real,
so, the imaginary part of HI potential, is usually treated
phenomenologically and its parameters are adjusted to optimize the
fit to the observed scattering. In the most cases, the Woods-Saxon
(\textit{WS}) shape (with volume or the surface type) is used for
the imaginary potential. Finally the HI potential can be written in
the general form as:
 \begin{eqnarray}
U(E,R)=N_{R}[U_{D}(E,R)+U_{EX}(E,R)]-iW_{V}[1+\exp (\dfrac{R-R_{V}}{a_{V}})]^{-1}  \nonumber\\
+4iW_{D}a_{D}\dfrac{d}{dR}[1+\exp (\dfrac{R-R_{D}}{a_{D}})]^{-1},
\end{eqnarray}
where the renormalization coefficient $ N_{R} $ together with the
parameters of the imaginary potential are adjusted to give the best
fit to the scattering data. The  renormalization coefficient $ N_{R}
$ is needed to account roughly for the many-nucleon exchange effects
and the dynamical polarization potential ($ \Delta U $) \cite{24}.
The volume or the surface $WS$ (the second and the third terms at
above formula) are usually used as the imaginary potential in the
elastic scattering analysis. However, we  only use the volume term
in our present calculations.

In the calculation of the exchange potential, we need also the
Coulomb potential, $ V_{C}(R) $. According to the reference
\cite{30}, the different models for the Coulomb potential do not
have serious effect on the theoretical predictions. So, in our
optical model (OM) calculations, we  chose the Coulomb potential to
be a simple interaction between a point charge and a uniform one
with the radius $ R_{C} $ \cite{2},
\begin{eqnarray}
V_{C}(R)=Z_{p}Z_{t}e^{2}\left\{
\begin{array}{cc}
 \dfrac{1}{R}& \;\;\;\; R > R_{C} \\\\

\dfrac{1}{2R_{C}}[3-(\dfrac{R}{R_{C}})^{2}]& \;\;\;\;R < R_{C}.
\end{array} \right.
\end{eqnarray}
with $ e^{2}=1.44$ $MeV.fm $ and $ R_{C}=R_{p}+R_{t} $, $
R_{i}=1.76Z_{i}^{1/3}-0.96 fm $, with $ i=p, t $.

\subsection{The choice of the effective interaction and the density distribution}
As it can be seen from the equations (2) and (3), the basic inputs
into the folding model are the nuclear densities of the colliding
nuclei in their ground state and the effective \textit{NN}
interaction. The density distributions  should be  normalized as:
\begin{equation}
\int\rho_{i}(\textbf{r}_{i})d{\bf{r}_{i}}=A_{i}
\end{equation}
where $ A_{i} $ is the mass number of the projectile or the target
nucleus. In this paper, the nuclear densities of two colliding
nuclei are approximated by the two-parameter Fermi distribution: $
\rho(r)=\rho_{0}[1+\exp ((r-c)/a)]^{-1} $ with parameters taken from
the  table 1 of the reference \cite{32}.

Given correct nuclear densities as inputs for the folding
calculations, it is still necessary to have an appropriate
\textit{NN} interaction for a reasonable prediction of the
nucleus-nucleus potential. The bare nucleon-nucleon interaction,
obtained from analysis of \textit{NN} scattering measurements, is
too strong to be used directly in the folding model, so, it is
common to use an effective in-medium interaction \cite{4,4p}. To
evaluate an in-medium \textit{NN} interaction starting from a
realistic free \textit{NN} interaction, still remains a challenge
for the nuclear many-body theory. Therefore, most of the microscopic
nuclear reaction calculations so far, still use different kinds of
effective in-medium \textit{NN} interaction \cite{3}. One of the
most popular choice for the \textit{NN} interactions, were based on
the \textit{M3Y} interactions and its density dependent versions
\cite{7,8,9,10,11,12,13}. These interactions are designed to
reproduce the \textit{G}-matrix elements of the \textit{Reid}
\cite{33} and the \textit{Paris} \cite{34}  \textit{NN} interactions
in an oscillator basis \cite{4,18,19,20} . We refer to these as the
\textit{M3Y-Reid} and the \textit{M3Y-Paris} interactions,
respectively. The explicit forms for the direct part of interactions
are \cite{4,4p}:
\begin{equation}
M3Y-Reid:
\;\;\upsilon_{D}(r)=[7999\dfrac{e^{-4r}}{4r}-2134\dfrac{e^{-2.5r}}{2.5r}]\:MeV,
\end{equation}
\begin{equation}
M3Y-Paris:
\;\;\upsilon_{D}(r)=[11062\dfrac{e^{-4r}}{4r}-2538\dfrac{e^{-2.5r}}{2.5r}]\:MeV
\end{equation}
whereas the exchange parts of interactions in the
finite-range-exchange ($FRE$) form (\textit{M3Y/FRE}) are written as
\cite{4,4p,2,3}:
\begin{equation}
M3Y-Reid:
\;\;\upsilon_{EX}(r)=[4631\dfrac{e^{-4r}}{4r}-1787\dfrac{e^{-2.5r}}{2.5r}-7.847\dfrac{e^{-0.7072r}}{0.7072r}]\:MeV,
\end{equation}
\begin{equation}
M3Y-Paris:
\;\;\upsilon_{EX}(r)=[-1524\dfrac{e^{-4r}}{4r}-518.8\dfrac{e^{-2.5r}}{2.5r}-7.847\dfrac{e^{-0.7072r}}{0.7072r}]\:MeV
\end{equation}
However, in many other calculations, the zero-range pseudo-potential
 (\textit{M3Y/PP}) is used to represent the knock-on exchange
 \cite{4,4p}. But in this work we focus on the finite range interactions i.e.
equations (13) and (15).

The older potentials based upon the density-independent \textit{M3Y}
interactions could reasonably reproduce the data of HI scattering at
the forward angle, or low energies \cite{4,4p}. Also, the
ground-state energy of nuclear matter (in a standard $Hartree-Fock$
calculation) using the \textit{M3Y} interactions is calculated in
the reference \cite{7}. One can realize that, the
density-independent M3Y interactions do not fulfill the saturation
condition for cold nuclear matter, i.e. leading to collapse. To
ensure the predication of the nuclear matter saturation, an
appropriate density-dependent factor is introduced into the original
\textit{M3Y} interaction. It is usually taken as an independent
factor that multiplied to the original radial \textit{M3Y}
interaction, i.e. $
\upsilon_{D(EX)}(r,\rho)=F(\rho)\upsilon_{D(EX)}(r) $. As it is
stated in the references \cite{4,4p}, there is no theoretical
justification for this factorization, but it leads to improve the
description of nuclear matter properties  and the HI scattering
data. Various forms for $ F(\rho) $ were proposed. In the $ DDM3Y1 $
and $ CDM3Yn \:(n=1-6) $, the following form is assumed for the
density dependent of the potential:
\begin{equation}
F(\rho)=C[1+\alpha \exp(-\beta\rho)-\gamma\rho]
\end{equation}
In $ BDM3Yn \:(n=0-3) $ interactions, a power-law dependent on $
\rho $ is supposed:
\begin{equation}
F(\rho)=C(1-\alpha\rho^{\beta})
\end{equation}
The parameters $ C $, $ \alpha $, $ \beta $ and $ \gamma $ are
adjusted to reproduce the saturation of cold symmetric nuclear
matter at $ \rho_{0}=0.17$ $ fm^{-3} $ and a binding energy per
nucleon of about $ 16$ $MeV $. The values of these parameters for $
CDM3Yn $ and $ DDM3Y1 $ and $ BDM3Yn $ interactions are given in the
references \cite{4,4p,7,30,35}. As we pointed out before for
comparison we focus on the finite range $DDM3Y1$ interaction
\cite{3}.

In the course of these application to the $ \mathcal{NN} $
scattering data, it is necessary to introduce an additional energy
dependent factor over which provided by localization of the exchange
potential:
\begin{equation}
\upsilon^{M3Y}_{D(EX)}(r,\rho,E)=\upsilon^{M3Y}_{D(EX)}(r)F(\rho)g(E)
\end{equation}
where $ g(E)=[1-k(E/A)] $ with $ k=0.002\:MeV^{-1} $ or $
k=0.003\:MeV^{-1} $ for the $Reid$ interaction or    the  $Paris$
interaction \cite{2}, respectively. However none of the above
potentials come from a Hamiltonian based many-body microscopic
calculations.

In the present work,  the $LOCV$ density dependent averaged
effective two-body interaction (\textit{AEI})   is generated though
the \textit{LOCV} method with the bare nucleon-nucleon
phenomenological \textit{Reid}68 potential, and  inserted  as an
input to the folding model calculations. In our previous work
\cite{14}, we obtained  the direct and the exchange parts of the
density dependent nucleon-nucleon \textit{AEI} as follows (see the
appendix for the definition of $a$ and $\mathcal{V}$):
\begin{equation}
\bar{\mathcal{V}}_{eff}^{D}(r,\rho)=\dfrac{\sum_{\alpha,i,j,k}(2T+1)(2J+1)\dfrac{1}{2}\mathcal{V}_{\alpha}^{j,k}(r,\rho)a_{\alpha}^{(i)^{2}}(r,\rho)}{\sum_{\alpha,i}(2T+1)(2J+1)\dfrac{1}{2}a_{\alpha}^{(i)^{2}}(r,\rho)},
\end{equation}
\begin{equation}
\bar{\mathcal{V}}_{eff}^{EX}(r,\rho)=\dfrac{\sum_{\alpha,i,j,k}(2T+1)(2J+1)\dfrac{1}{2}[(-1)^{L+S+T}]\mathcal{V}_{\alpha}^{j,k}(r,\rho)a_{\alpha}^{(i)^{2}}(r,\rho)}{\sum_{\alpha,i}(2T+1)(2J+1)\dfrac{1}{2}[(-1)^{L+S+T}]a_{\alpha}^{(i)^{2}}(r,\rho)},
\end{equation}
where $ \alpha=JLST $, $J$ is the total orbital angular momentum of
two nucleons i.e. $ L $ plus  $ S $, and $ T $,  is the total
iso-spin of two nucleons. Then we have reformulated these
interactions as the product of a pure radial and a pure
density-dependent parts:
\begin{equation}
\bar{\mathcal{V}}_{eff}^{D(EX)}(r,\rho)=\bar{\mathcal{V}}^{D(EX)}(r)F^{D(EX)}(\rho).
\end{equation}
Here, we  chose $ \bar{\mathcal{V}}^{D(EX)}(r) $ and $
F^{D(EX)}(\rho) $ to give the best fit to the $LOCV$
$\bar{\mathcal{V}}_{eff}^{D(EX)}(r,\rho) $ and the corresponding
equation of state ($LOCV$-$EOS$) of nuclear matter. The reader
should note that, by this statement, we mean that the fitted
potentials should again reproduce the $SNM$ saturation properties
given in the table 1.

There are many different functions which can fit $ F^{D(EX)}(\rho) $
well enough. A detailed role of description of density-dependent
factor ($F$) can be found in our previous work, the reference
\cite{14}, where we   stated that the $LOCV AEI$ includes a radial
part and a density-dependent part and we   show that, the radial
part form of the $LOCV AEI$ is fixed in any density (exactly like
the M3Y type interactions) and the $EOS$ of $SNM$ without taking
into account the density-dependent factor did not fulfill the
saturation condition and the system was collapsed (see the figure 7
of the reference \cite{14}). But one should notice that our
density-dependent factor is not an external factor and it comes from
the $LOCV$ calculations. So, we   just parameterized it in a
suitable form (i.e. see below, the equation (23)) (the exponential
dependent form for density). In the reference [21], we   compared
the direct and exchange parts of the $LOCV AEI$ with the
corresponding results of the M3Y interactions.(see the figures (1)
and (4) of the reference \cite{14})

So as we stated above, similar to our previous work \cite{14}, in
order to reproduce the $LOCV$-$EOS$ of nuclear matter properly, we
use the power-law-dependent on $ \rho $: $
F^{D(EX)}(\rho)=\mathcal{C}^{D(EX)}(1-\alpha^{D(EX)}\rho^{\beta^{D(EX)}})
$. In this paper, we use the exponential dependent form for $ \rho $
(similar to the  $ DDM3Y1 $ interaction):
\begin{equation}
F^{D(EX)}(\rho)=\mathcal{C}^{D(EX)}(1+\alpha^{D(EX)}\exp
(-\beta^{D(EX)}\rho)).
\end{equation}
This choice allows us to easily calculate  the integration of the
double-folding equations in the momentum space \cite{4,4p}. The
parameters of equation (23) are given in the  table 2.

Similar to the \textit{M3Y} interactions, in order to apply the
\textit{LOCV AEI} to the $ \mathcal{NN} $ scattering data, we need
to add an explicit energy-dependent factor to our \textit{LOCV AEI}
to obtain the best description of HI scattering by taking into
account the variation in the incident energy. We   found that this
factor can be assumed as the linear dependent to the incident energy
per nucleon, which is similar to the \textit{M3Y} interactions i.e.
$ g(E)=[1-k(E/A)] $. So, we can rewrite the \textit{LOCV AEI} as:
\begin{equation}
\bar{\mathcal{V}}_{eff}^{D(EX)}(r,\rho,E)=\bar{\mathcal{V}}^{D(EX)}(r)F^{D(EX)}(\rho)g(E).
\end{equation}
Here, as in other $HI$ works,  the $ k $ is chosen to give the best
fit to the $ \mathcal{NN} $ scattering data. It is shown that in the
case of our \textit{LOCV AEI} by choosing $ k=0.003 MeV^{-1} $, the
optimized fit will be acquired. However, the calculation is not very
sensitive to this parameter if  it is  chosen in  its order.
\subsection{The Computational procedure }
At first, we  calculate the real part of the folded potential for $
^{12}C-^{12}C $ and $ ^{16}O-^{16}O $ elastic scattering by the
double folding formula, i.e. the equations (2) and (3). Then  we use
the \textit{LOCV AEI} as the effective \textit{NN} interactions and
the two-parameter Fermi distribution for the nuclear densities of
the projectile and the target nuclei. Now, in order to compute the
 scattering differential cross section,
  we also use the $FRESCO$ code developed by Ian Thompson
\cite{36p} which  is developed for the calculation of different
types of nucleon-nucleus and nucleus-nucleus scattering
cross-sections. This code is capable to use our folded potential
directly, to calculate the elastic scattering cross section.

We will discuss our resulting potentials and the elastic  scattering
cross section for $^{12}C-^{12}C$ and $^{16}O-^{16}O$ systems in the
next section. Generally, the goodness of our resulting  cross
section is quantified via the $ \chi^{2} $ expression \cite{4,4p},
 \begin{equation}
\chi^{2}=\dfrac{1}{N_{\sigma}}\sum_{i=1}^{N_{\sigma}}\dfrac{(\sigma_{th}-\sigma_{ex})^{2}}{(\Delta\sigma_{ex})^{2}}
\end{equation}
where $ \sigma_{th} $ and  $ \sigma_{ex} $  are the theoretical and
the experimental cross sections and  $ \Delta\sigma_{ex} $ are
defined as the uncertainties in the experimental cross sections,
respectively. $ N_{\sigma} $ is the total number of angles at which
measurements are made.
\section{RESULTS AND DISCUSSIONS}
As it was pointed out in the previous section, in order to calculate
the direct and the exchange components of the real part of the HI
optical potential, we  use the direct and the exchange parts of the
\textit{LOCV AEI} as the effective \textit{NN} potential in the
double folding formula (the equations (2) and (3)). Since the wave
number of relative motion $ k_{rel}(R) $, the equation (4), depends
on the total HI potential, we are faced with a self-consistency
problem in obtaining the exchange part of the HI potential at each
radial point. So, we  apply  the iterative method at each point and
use $ U_{D}(E,R) $ as the starting potential to enter $
\hat{j}_{0}(k(\textbf{R})s/M) $ in the exchange integral, the
equation (9), i.e. as it is performed when one considers the
\textit{M3Y} interactions in the folding formula \cite{3}.

Unfortunately at small internuclear distances ($ R\leq 1 fm $), the
iterative method for calculating the exchange potential based on the
\textit{LOCV AEI},  does not converge reasonably. Of course, with
increasing the incident energy, this problem will be solved.  Due to
this  low convergence speed of iterative method in case of the
insertion of  the \textit{LOCV AEI} in the folding formula, we need
much more number of iterations with respect to the \textit{M3Y}
interactions, in  obtaining  the exact self-consistent results for $
U_{EX}(E,R) $, especially at small internuclear distances. According
to the reference \cite{3}, in the case of the \textit{M3Y}
interactions, the number of iterations required is around 20 at
smallest radii and ranges from 3 to 5 at the surface region, while,
in case of the \textit{LOCV AEI}, it is around 150 to 200 at
smallest radii and around 2 or 3 at the surface region. For this
reason, too much CPU computer time is needed to calculate the
exchange part of the HI potential in case of  the \textit{LOCV AEI}.
For example for the $^{12}C-^{12}C $ elastic scattering at the $
E_{lab}=300 MeV $, it took about 50 hours computer CPU time by using
the high performance computing (HPC) machine of the university of
Tehran. Because of the different radial shapes of the \textit{LOCV
AEI} with respect to the \textit{M3Y} interactions at the small
distances, this problem is expected. Conversely to the \textit{M3Y}
potentials, due to short range correlations coming from the
channel-dependent correlation functions, at very small distances,
the direct and the exchange components of the \textit{LOCV AEI} go
to zero (see the figures 1 to 4 of the reference \cite{14}) and this
behavior makes the iterative method not to converge at these
distances as quicker as for the \textit{M3Y} interactions. While,
since the \textit{M3Y} interactions are constructed from the
selected channels of, for example the \textit{Reid}68 potential,
i.e. the singlet and the triplet even and odd components, one does
not faced with this problem.

So in the figures 1 and 2, we   plot the calculated    direct,
 exchange   and also the total components of the folded potential  by using
the $LOCV$ $AEI$ for $ ^{12}C-^{12}C $ and $^{16}O-^{16}O$ systems
at several incident energies i.e. 112, 126.7, 240, 300 and 360 $MeV$
for $ ^{12}C- ^{12}C $ and 124, 145, 250, 350 and 480 $MeV$ in the
case of $ ^{16}O- ^{16}O $ (note that we extrapolate the folded
potential at the small distances ($R<1fm$)  for some points that the
iterative method is not converge rapidly for calculation of the
exchange potential based on $LOCV-AEI$). Comparing the exchange
parts with the direct parts at each incident energy, one can observe
that the most of energy dependence of the HI potential is arising
from the exchange part, as one should expects. We  also notice that
at small internuclear distances, which corresponds to large overlap
densities ($ \rho > \rho_{0} $), the exchange potential is more deep
than the direct potential, especially at lower energies, and this
shows that the density-dependent contribution of HI potential
predominately comes from the exchange term. On the other hand,  in
the surface region, which corresponds to the small overlap
densities, all the calculated direct and exchange potentials are
close in the strength and the slope. The figures 1 and 2 also show
that with increasing the incident energy of projectile, the depth of
the HI potential at the origin is decreased systematically. Similar
results
  already   reported in calculating folded potential using the
\textit{M3Y} interactions, for example see the references
\cite{3,5}.

We compare our calculated folded potential, using the $LOCV AEI$
with the corresponding results of DDM3Y1 \cite{3} for the cases of
the $^{12}C-^{12}C$ at $E_{lab}=300$ $MeV$ and the $^{12}O-^{12}O$
at $E_{lab}=350$ $MeV$ in the figures 3 and 4, respectively. It can
be observed that the folded potentials by using the $LOCV AEI$ are
more deep than the DDM3Y1 ones.  For the other energies, the similar
results are obtained.

The results of our folding analysis for the $^{12}C-^{12}C$ elastic
scattering, at incident energies ranging from 112 to 360 $MeV$ with
  FRESCO code are presented in the figure 5 while the table 3
shows  the $WS$ parameters of the imaginary part of HI potential for
the same system and at the same energies as well as $\sigma_R$ and
$\chi^2$ (with respect to the experimental data, see the next
paragraph). In this paper we  take the imaginary part of HI
potential as the conventional $WS$ form and adjust its parameters to
obtain the best description of the experimental scattering data in
the whole angular range at each incident energy. The parameters  in
the table 3  are close to those found in earlier analysis for $
DDM3Y1-Reid $ (see the table 2 of the reference \cite{3}). The table
3 also shows that the best fit to the scattering data, can be found
by using the values of $ N_{R} $ which are slightly deviated from
the unity. This result indicates that the high-order effects are
negligible in our calculations.

The different panels of figure 5 (a to e) show the calculated cross
section of $ ^{12}C- ^{12}C $ elastic scattering at several incident
energies, i.e. 112, 126.7, 240, 300 and 360 $MeV$, by using the
\textit{LOCV AEI} folded potential in the   FRESCO code. The
scattering experimental data \cite{37,38,39,40,41,42,43,44,45} and
the resulting cross sections of the \textit{DDM3Y1} \cite{3} are
also presented. It is observed that a quite good description of data
scattering can be obtained by using the \textit{LOCV AEI} and
adjusting the imaginary potential parameters and renormalization
coefficient. However, in comparison to the \textit{DDM3Y1} ($Reid$)
results \cite{3}, our results may not be too satisfactory,
especially at forward angles, but one should notice that
\textit{DDM3Y1} potential was constructed from the selected channels
of the \textit{Reid68} potential and its density dependent factor
was added to it later,  to provide a reasonable description of HI
scattering data and the equation of state ($EOS$) of nuclear matter,
while the \textit{LOCV AEI} are constructed based on the many-body
calculations without any free parameters in the $LOCV$ calculations
and its density dependent part comes directly from the \textit{LOCV}
formalism (obviously $LOCV$ formalism has its owns $EOS$, i.e.
$LOCV$-$EOS$). It is worth to say that, by increasing the incident
energy a better fit to the  scattering data is achieved  using the
\textit{LOCV AEI} at forward angles.

The calculated cross sections using the \textit{LOCV AEI} for $
^{16}O-^{16}O$ elastic scattering at incident energies ranging from
124 to 480 MeV are plotted in the different panels (a to e) of the
figure 6. The scattering experimental data
\cite{37,38,39,40,41,42,43,44,45} show a clear refractive pattern at
large angles and a diffractive pattern produced by an interference
between nearside and farsight components of the scattering amplitude
at the small angles. The refractive pattern can be clearly
distinguished from the diffractive structure, i.e. it is shifting
substantially    towards the small angles with increasing the
incident energy \cite{5}.

One can realize that our calculated cross sections can predict
reasonably the behavior of  scattering data  on large ranges of
scattering angles \cite{37,38,39,40,41,42,43,44,45}. Similar to the
results obtained above for $ ^{12}C-^{12}C $ system, there exist
considerable differences between our results with respect  to the
experimental data  and those coming from \textit{DDM3Y1}. Again, the
similar discussion can be made for these results as the one we made
above for $^{12}C$. In this case, it can also be observed that the
agreement of our calculations to the scattering data
 are getting better as the energies of projectile
are increased. To improve the agreement of the calculated cross
sections using the \textit{DDM3Y1-Reid} and \textit{DDM3Y1-Paris}
with data in the large-angle region, in the references \cite{3,5} a
surface ($WSD$) term was included into the imaginary part of
potential. We hope, in our future works, we could investigate the
inclusion of the $WSD$ term for improving our results.

The table 4 shows the parameters of our $WS$ imaginary potential and
renormalization coefficient for $ ^{16}O-^{16}O $ system at
different incident energies as above. Again, we can see the values
of $ N_{R} $ are close to the unity and our $WS$ parameters are in
agreement to the $WS$ parameters of \textit{DDM3Y1} analysis
\cite{3}.
\section{SUMMARY}
In conclusion, we   analyzed the experimental data of $ ^{12}C-
^{12}C $ and $ ^{16}O- ^{16}O $ elastic scattering at different
incident energies, within the standard optical model ($OM$), using
the density-dependent \textit{LOCV AEI}. The direct and the exchange
parts of \textit{LOCV AEI} were generated based on the \textit{LOCV}
method for the symmetric nuclear matter, using  the \textit{Reid}68
interaction as the input phenomenological potential. In order to use
our interaction into the folding model, we   separated the radial
and the density-dependent parts of the \textit{LOCV AEI}. Our
calculated cross sections for $ ^{12}C $ and $ ^{16}O $ systems,
indicate that a quite reasonable description of data scattering can
be obtained by using the \textit{LOCV AEI} and adjusting the
imaginary potential parameters and the  renormalization coefficient.
Our calculations favor a rather weak imaginary potential and a small
deviation of the renormalization factor from the unity. Comparing
our calculations with corresponding results of the \textit{DDM3Y1},
show some considerable differences. But one should notice that the
\textit{M3Y} interactions are semi-phenomenological potentials and
they are  constructed from the selected channels of the \textit{Reid
} potential, i.e. the singlet and the triplet even and odd
components and the parameters of its density dependent part are
adjusted to gain a reasonable description of  HI scattering data and
the EOS of nuclear matter. So, it is natural to  fit the scattering
data better than  ours. While the  \textit{LOCV AEI} are based on
the many-body calculation with the phenomenological \textit{NN}
potential without any free parameters, i.e. there are no free
parameters in the $LOCV$ formalism besides the \textit{NN} potential
and its density dependent part comes directly from the self
consistent \textit{LOCV} calculations. So
 it is meaningful to apply the $LOCV AEI$ interaction to the
heavy-ion scattering as the first attempt, but we hope  the
improvement of the present model could  be committed in the near
future.

The spite of the slow  convergence speed of iterative procedure in
using the \textit{LOCV AEI} in calculating of the exchange
potential, especially at small internuclear distances which
increases the computing  time, since the \textit{LOCV AEI} are based
on the many-body calculations, they are more trustable for the $
\mathcal{NN} $ collision calculations. So, with respect to the above
arguments, because the \textit{LOCV AEI} provides a reasonable
description of the normal nuclear matter \cite{14} as well as the HI
elastic scattering  data simultaneously, we can claim the
\textit{LOCV AEI} is a good candidate to approximate the \textit{NN}
interaction for the   nuclear matter and finite nuclei .

Finally we should make this comment that the insertion of other
 phenomenological nucleon-nucleon potential such as the $Av_{18}$
potential, should not have any dramatic change on our present
results, but it is worth to be investigated.
\begin{acknowledgements}
The authors would like  to acknowledge the Research Council of
University of Tehran and the Iran National Science Foundation
($INSF$) for the grants provided for them. They also would like to
sincerely thank  Professor Ian Thompson for his valuable help
regarding the FRESCO code.
\end{acknowledgements}

\newpage
\begin{table}
\caption{ The nuclear matter saturation parameters (for
\textit{Reid} and $\triangle$-\textit{Reid} potentials) extracted
from
 reference \cite{464}($ E_{3} $   denotes the
inclusion of the three-body cluster energy, see the appendix A).}
\begin{center}
\begin{tabular}{c c|c|c c|c| c}
\hline
  & \multicolumn{2}{c}{With Reid} &  & \multicolumn{2}{c}{With $\triangle$-Reid}  &   \\  \cline{2-3} \cline{5-6}

   & $ E_{1}+E_{2} $ & $ E_{1}+E_{2}+E_{3} $ & & $ E_{1}+E_{2} $ & $ E_{1}+E_{2}+E_{3} $ & Empirical  \\
\hline

Saturation Fermi  & 1.61 & 1.46 & & 1.55 & 1.44 & 1.38  \\
  momentum $ (fm^{-1}) $ &   &  &  &  &   &    \\

 Saturation binding & 22.54 & 21.85 & & 16.28 & 15.52 & 15.86 \\
 energy $ (MeV) $ &   &   & &   &   &   \\

Compressibility $ (MeV) $ & 340 & 298 & & 300 & 277 & (200-300) \\

 Convergence parameter & 0.127 & 0.085 & & 0.093 & 0.062 &    \\
\hline
\end{tabular}
\end{center}
\label{1}
\end{table}
\begin{table}[H]
\caption{The parameters of the density-dependent part of the direct
and the exchange components ($ F^{D(EX)}(\rho) $) of the $LOCV$
$AEI$ using the  \textit{Reid}68 interaction as the input
potentials.}
\begin{center}
\begin{tabular}{c|c c c }
 \hline        & $ \mathcal{C} $ & $ \alpha $ & $ \beta $ \\
  \hline Direct component & 0.38 & 5.03 & 3.22 \\
  Exchange component & 13.57 & -0.9 & 0.12 \\
\hline
\end{tabular}
\end{center}
\end{table}
\newpage
\begin{table}[H]
\caption{The $WS$ parameters of the imaginary part of HI potential
used in our   folding analysis of the $ ^{12}C-^{12}C  $ elastic
scattering at $ E_{lab}=112, 126.7, 240, 300, 360$ $ MeV $.}
\begin{center}
\begin{tabular}{c|c c c c c c }
\hline
\hline \\
 $ E_{lab}(MeV) $ & $ N_{R} $ & $ W_{V}(MeV) $ & $ R_{V}(fm) $ & $ a_{V}(fm) $ & $ \sigma_{R}(mb) $ & $ \chi^{2} $ \\
\\
\hline\\
 112 & 0.9383 & 17.4 & 5.403 & 0.70 & 1526.79 & 36.52 \\
\\
126.7 & 0.9230 & 19.10 & 5.128 & 0.79 & 1563.51 & 41.86 \\
 \\
 240 & 1.0207 & 28.90 & 5.266 & 0.69 & 1551.95 & 39.34 \\
\\
300 & 0.9731 & 33.82 & 4.991 & 0.72 & 1497.85 & 18.33 \\
  \\
 360 & 0.9684 & 34.5 & 4.808 & 0.70 & 1374.73 & 9.81 \\
\hline \hline
\end{tabular}
\end{center}
\end{table}
\newpage
\begin{table}[H]
\caption{The same as the table 5 but for the $ ^{16}O- ^{16}O $
elastic scattering at $ E_{lab}=124, 145, 250, 350, 480$ $ MeV $. }
\begin{center}
 \begin{tabular}{c|c c c c c c }
\hline
\hline \\
 $ E_{lab}(MeV) $ & $ N_{R} $ & $ W_{V}(MeV) $ & $ R_{V}(fm) $ & $ a_{V}(fm) $ & $ \sigma_{R}(mb) $ & $ \chi^{2} $ \\
\\
\hline\\
 124 & 0.9455 & 15.3 & 6.30 & 0.93 & 2201.99 & 34.34 \\
\\
145 & 1.007 & 16.4 & 6.199 & 0.95 & 2226.17 & 37.07 \\
 \\
 250 & 1.011 & 31.6 & 5.695 & 0.86 & 2091.89 & 39.71 \\
\\
350 & 0.9890 & 36.76 & 5.544 & 0.77 & 1876.58 & 21.19 \\
  \\
 480 & 0.9703 & 42.65 & 5.241 & 0.79 & 1778.03 & 42.37 \\
\hline \hline
\end{tabular}
\end{center}
\end{table}
\newpage
\newpage
\begin{figure}[h!]
\includegraphics [scale=0.6]{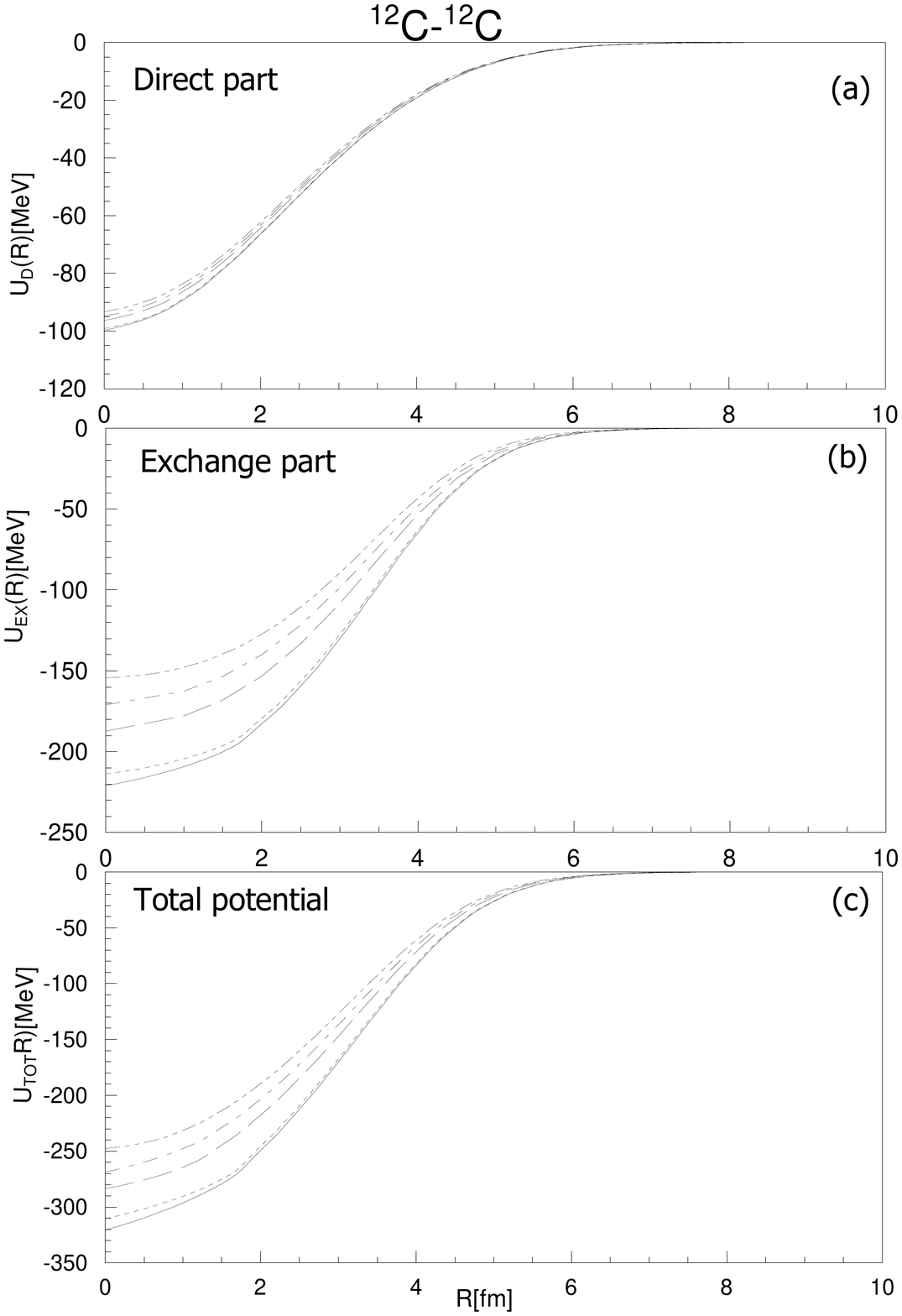}
\caption{The calculated direct and the exchange components and the
total folded potential, by using $LOCV$ $AEI$ for the $ ^{12}C-
^{12}C $ system at the several incident energies, i.e. $E_{lab}$
=112 (the full curve), 126.7 (the short-dash curve), 240 (the
long-dash curve), 300 (the long-short-dash curve), 360
(long-double-short-dash curve) $MeV $.}
\end{figure}
\begin{figure}[h!]
\includegraphics [scale=0.6]{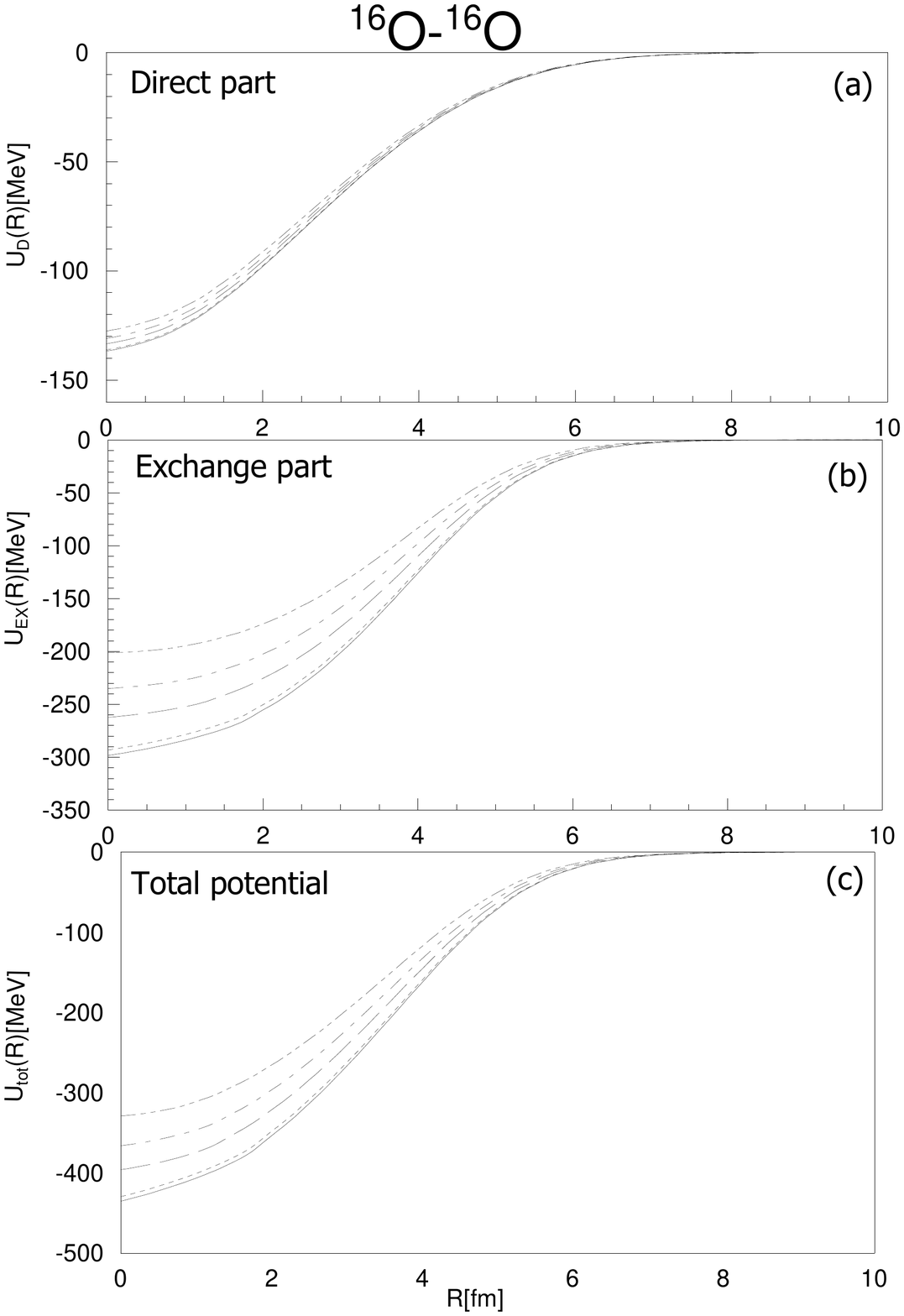}
\caption{As the figure 1 but for the $ ^{16}O- ^{16}O $ system and
$E_{lab}$=124 (the full curve), 145 (the short-dash curve), 250 (the
long-dash curve), 350 (the long-short-dash curve), 480 (the
long-double-short-dash curve) $MeV $.}
\end{figure}
\begin{figure}[h!]
\includegraphics [scale=0.6]{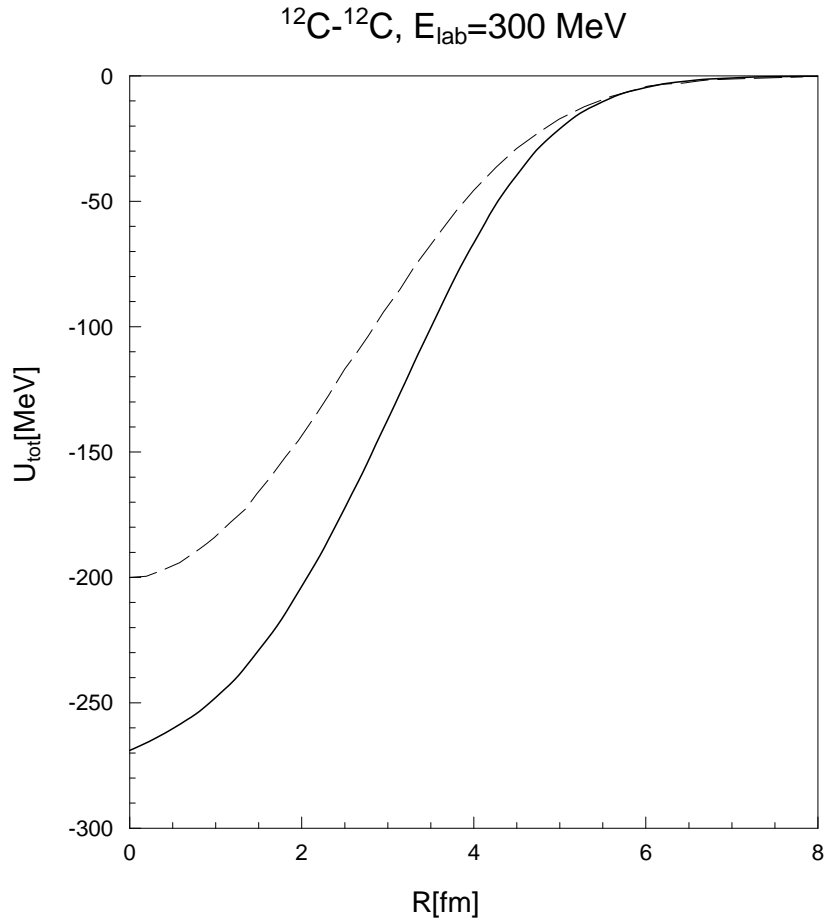}
\caption{The comparison of the calculated folded potentials using
the \textit{LOCV AEI} (the full curve) and the DDM3Y1 \cite{3} (the
short-dash curve) potential for the $ ^{12}C- ^{12}C $ scattering at
$ E_{lab}=300 $ $MeV $.}
\end{figure}
\begin{figure}[h!]
\includegraphics [scale=0.6]{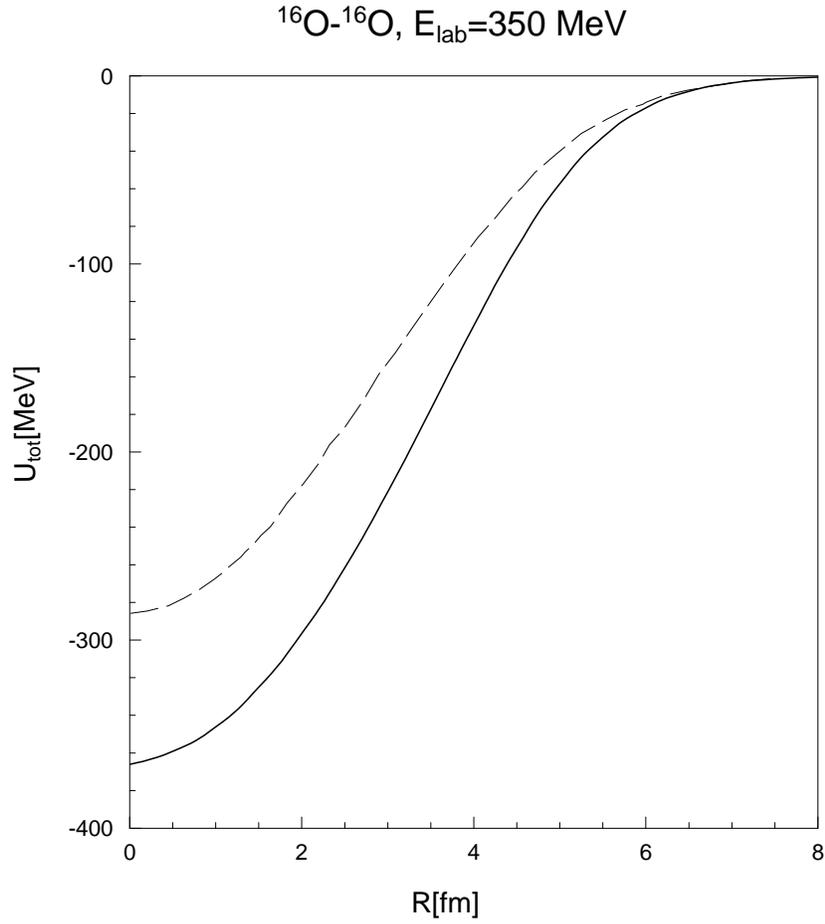}
\caption{As the figure 3 but for the $ ^{16}O-^{16}O $ scattering at
$ E_{lab}=350$ $MeV $.}
\end{figure}
\begin{figure}[h!]
\includegraphics [scale=0.6]{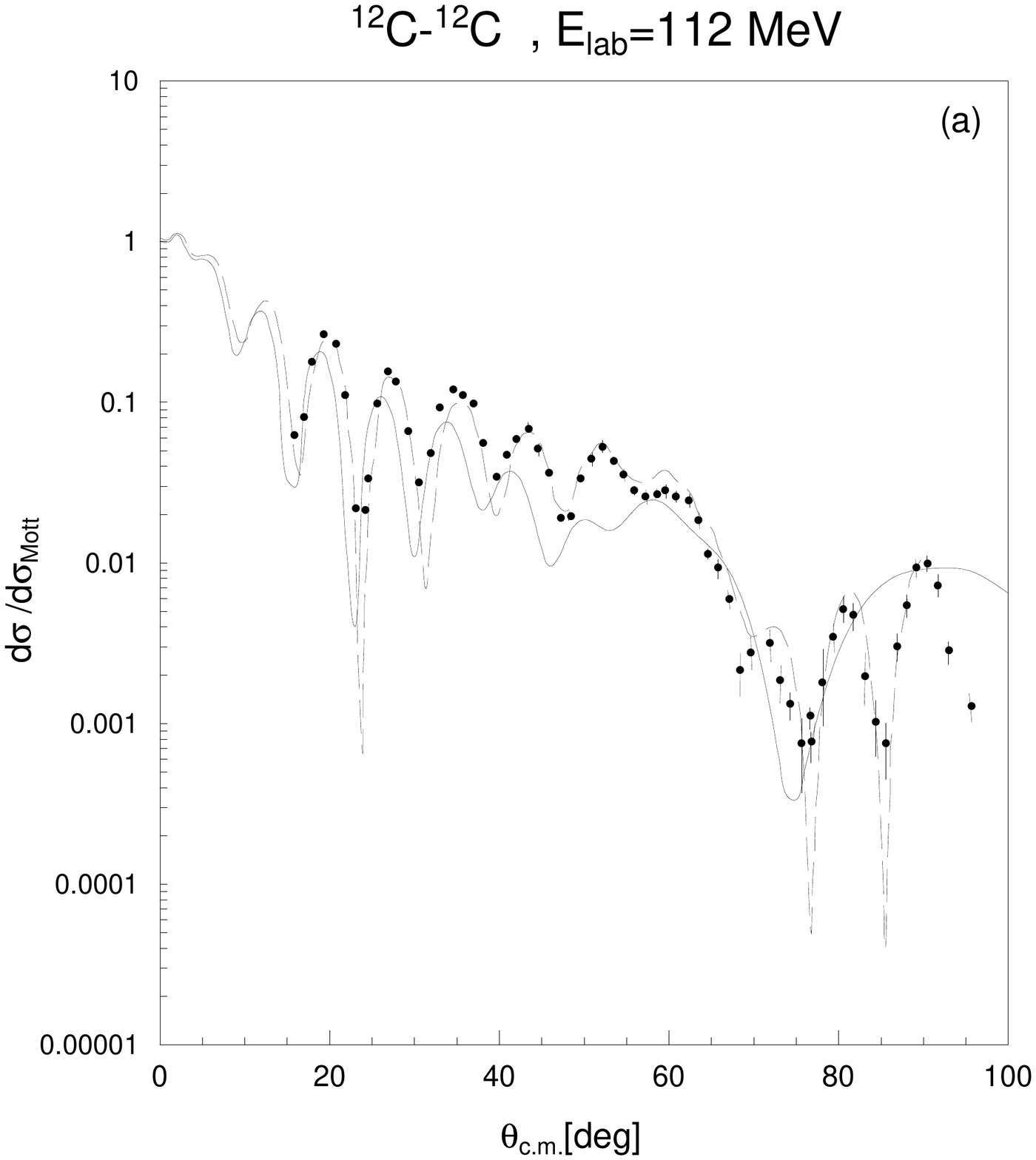}
\end{figure}
\begin{figure}[h!]
\includegraphics [scale=0.6]{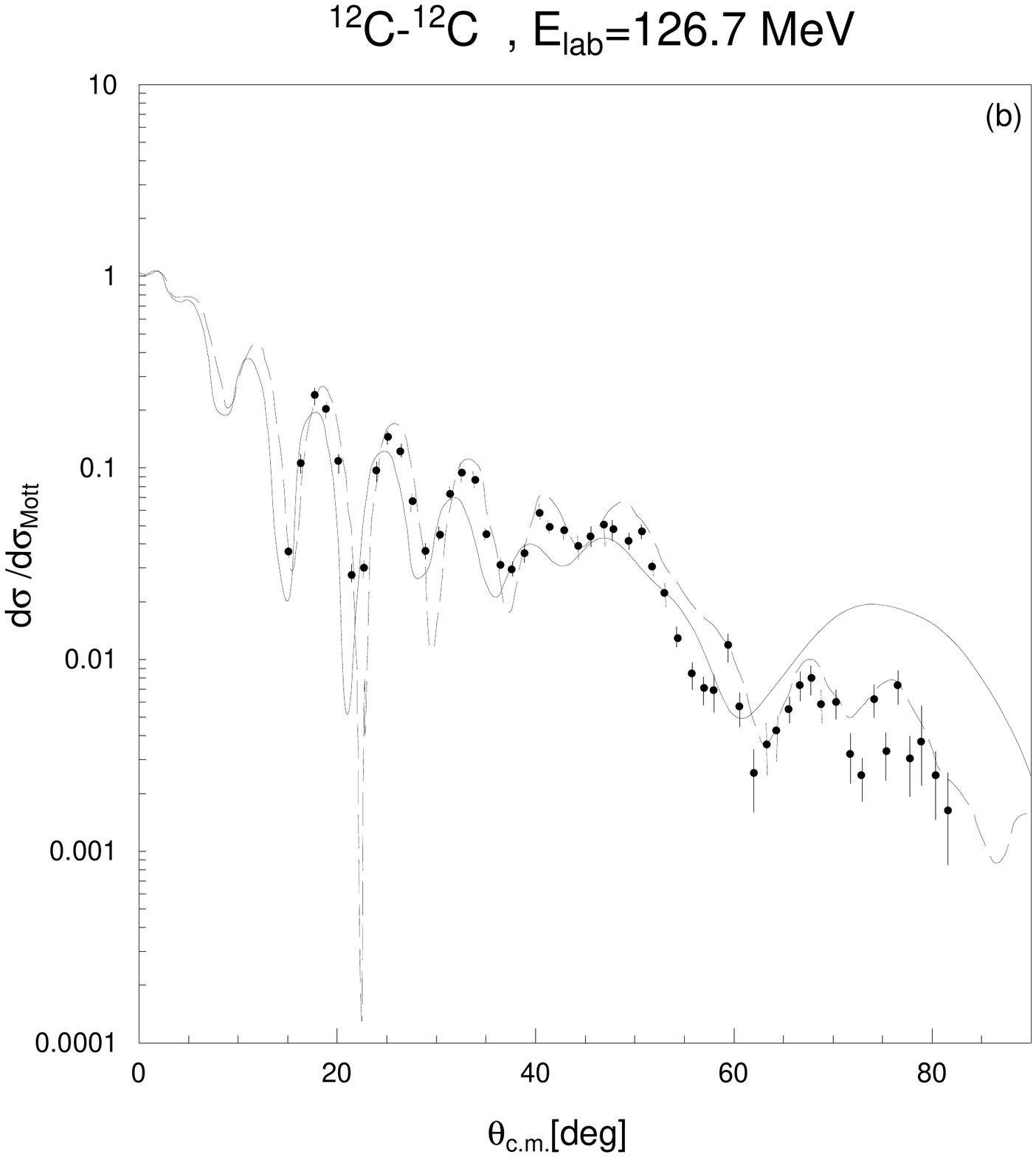}
\end{figure}
\begin{figure}[h!]
\includegraphics [scale=0.6]{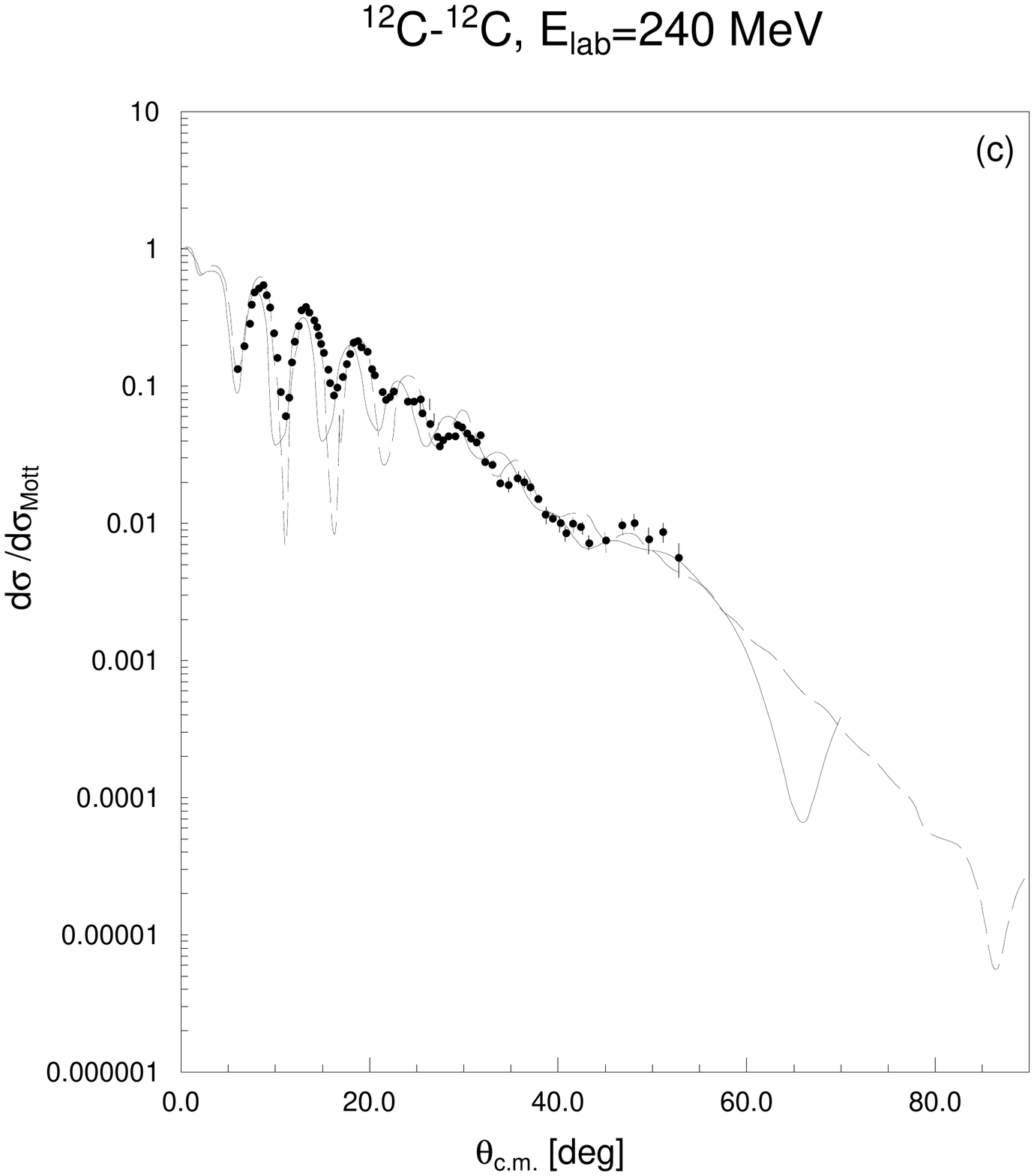}
\end{figure}
\begin{figure}[h!]
\includegraphics [scale=0.6]{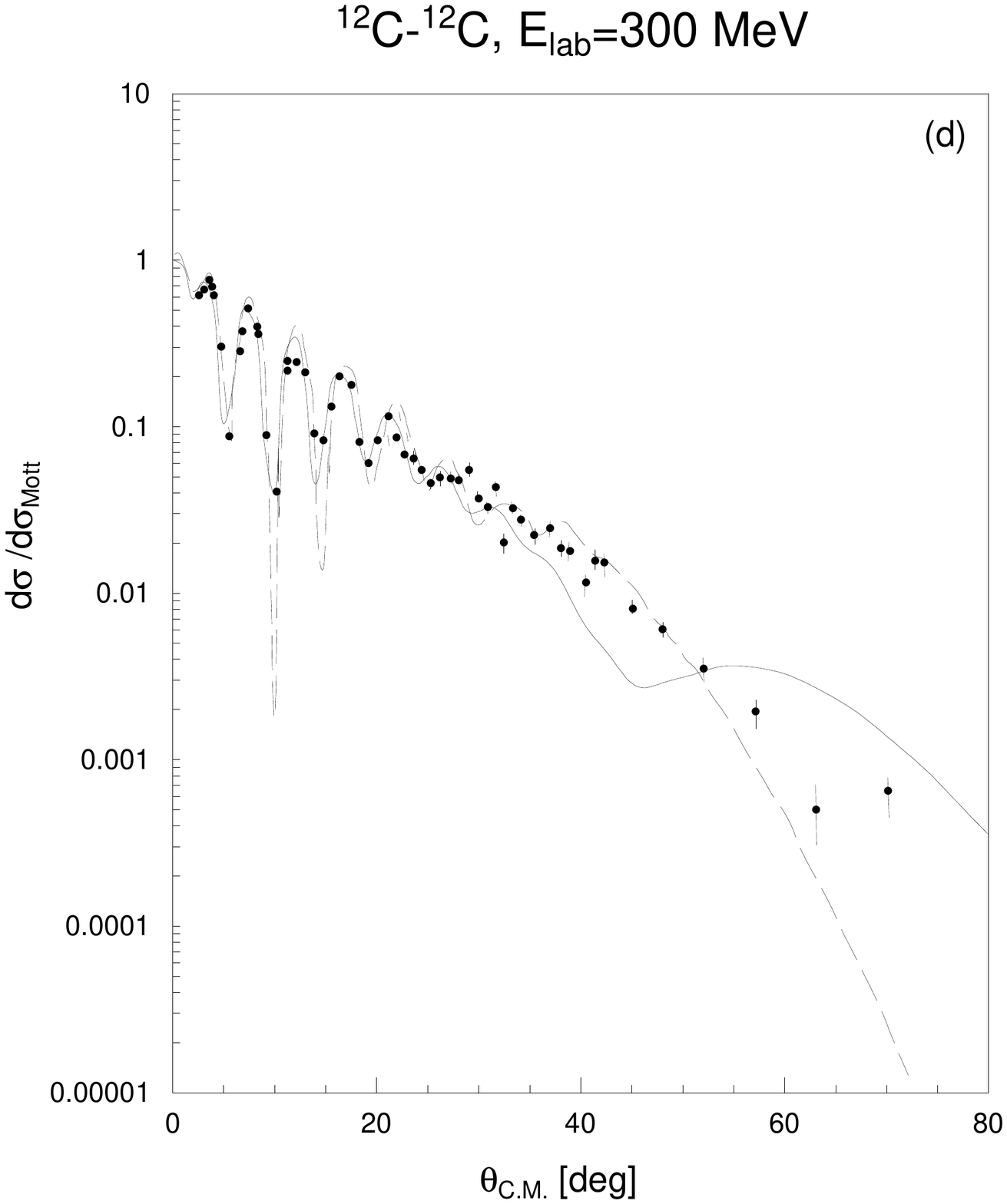}
\end{figure}
\begin{figure}[h!]
\includegraphics [scale=0.6]{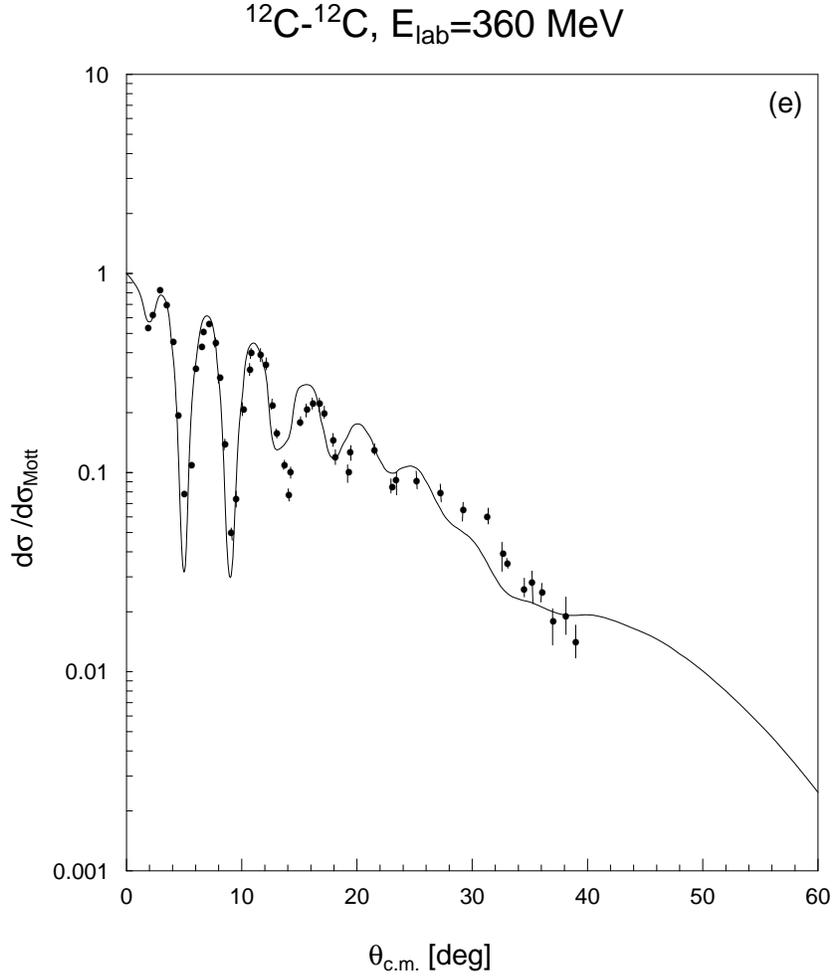}
\caption{The calculated cross sections of the $^{12}C-^{12}C $
elastic scattering at $ E_{lab}=112, 126.7, 240, 300, 360$ $MeV $ by
using the \textit{LOCV AEI} (the full curve) using the  FRESCO code.
The experimental scattering data (the full dotted points) and the
resulting cross section of the finite range interaction  DDM3Y1
\cite{3} (the dash curve) are also presented. The experimental data
are taken from the references \cite{37,38,39,40}.}
\end{figure}
\begin{figure}[h!]
\includegraphics [scale=0.6]{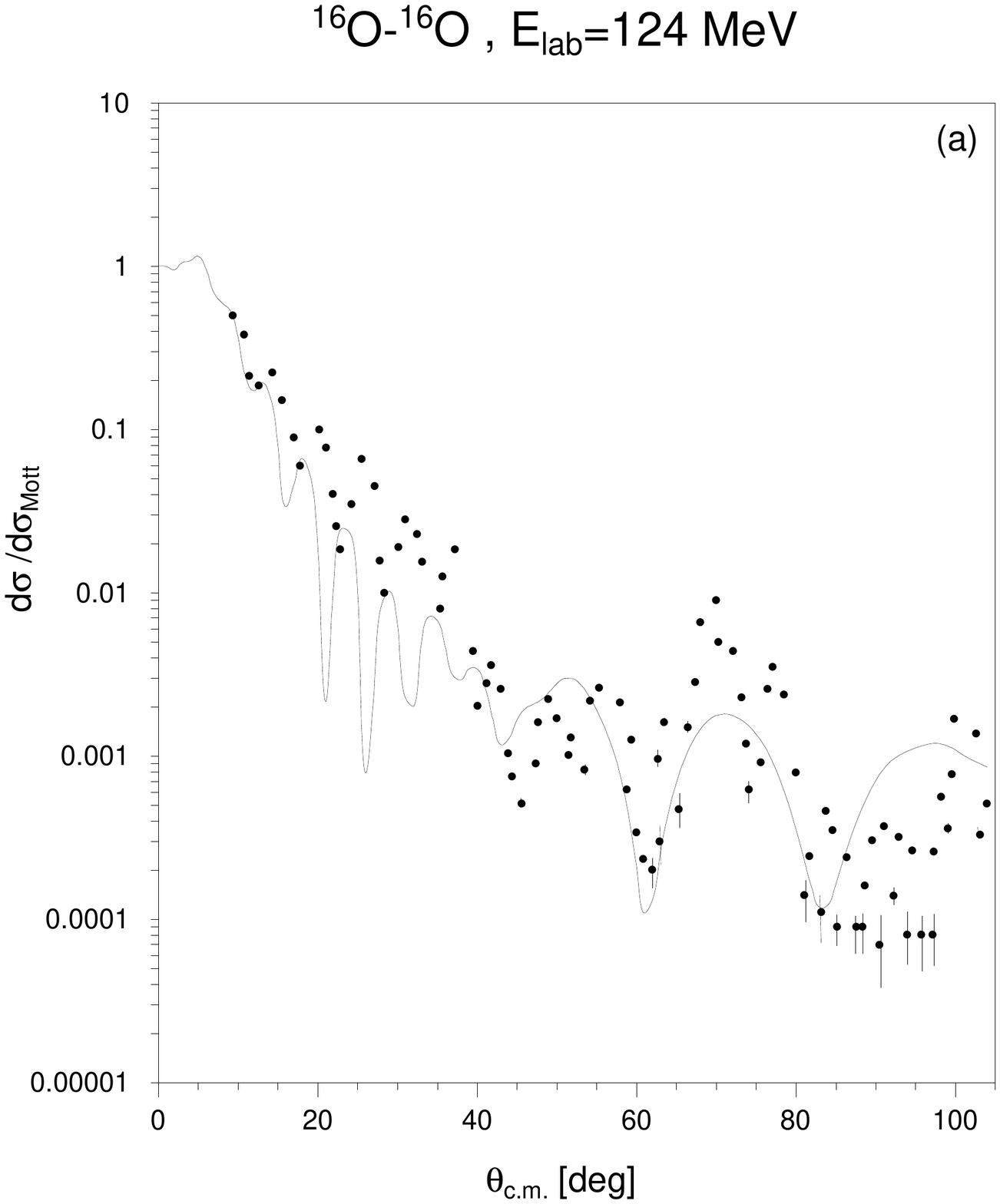}
\end{figure}
\begin{figure}[h!]
\includegraphics [scale=0.6]{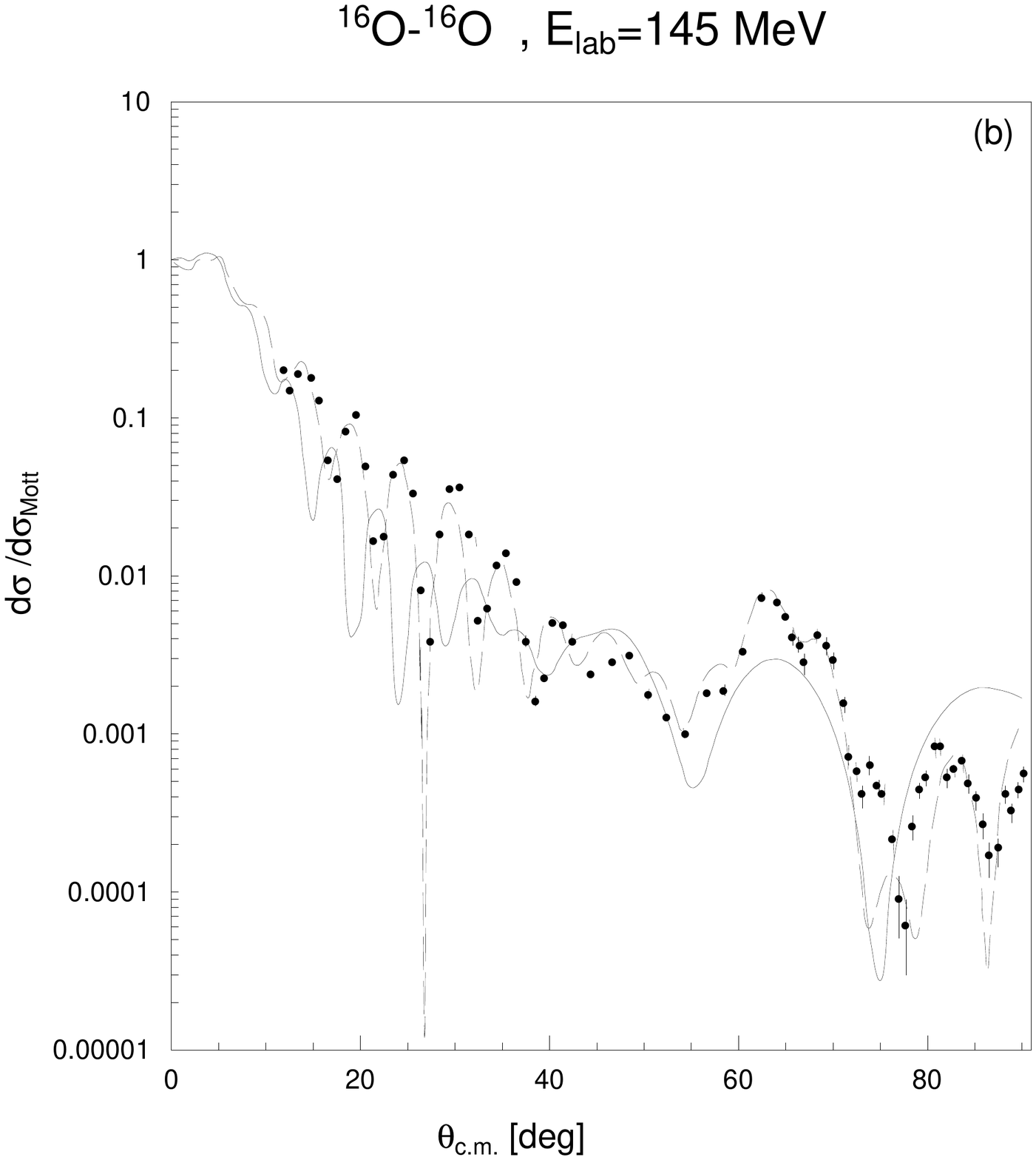}
\end{figure}
\begin{figure}[h!]
\includegraphics [scale=0.6]{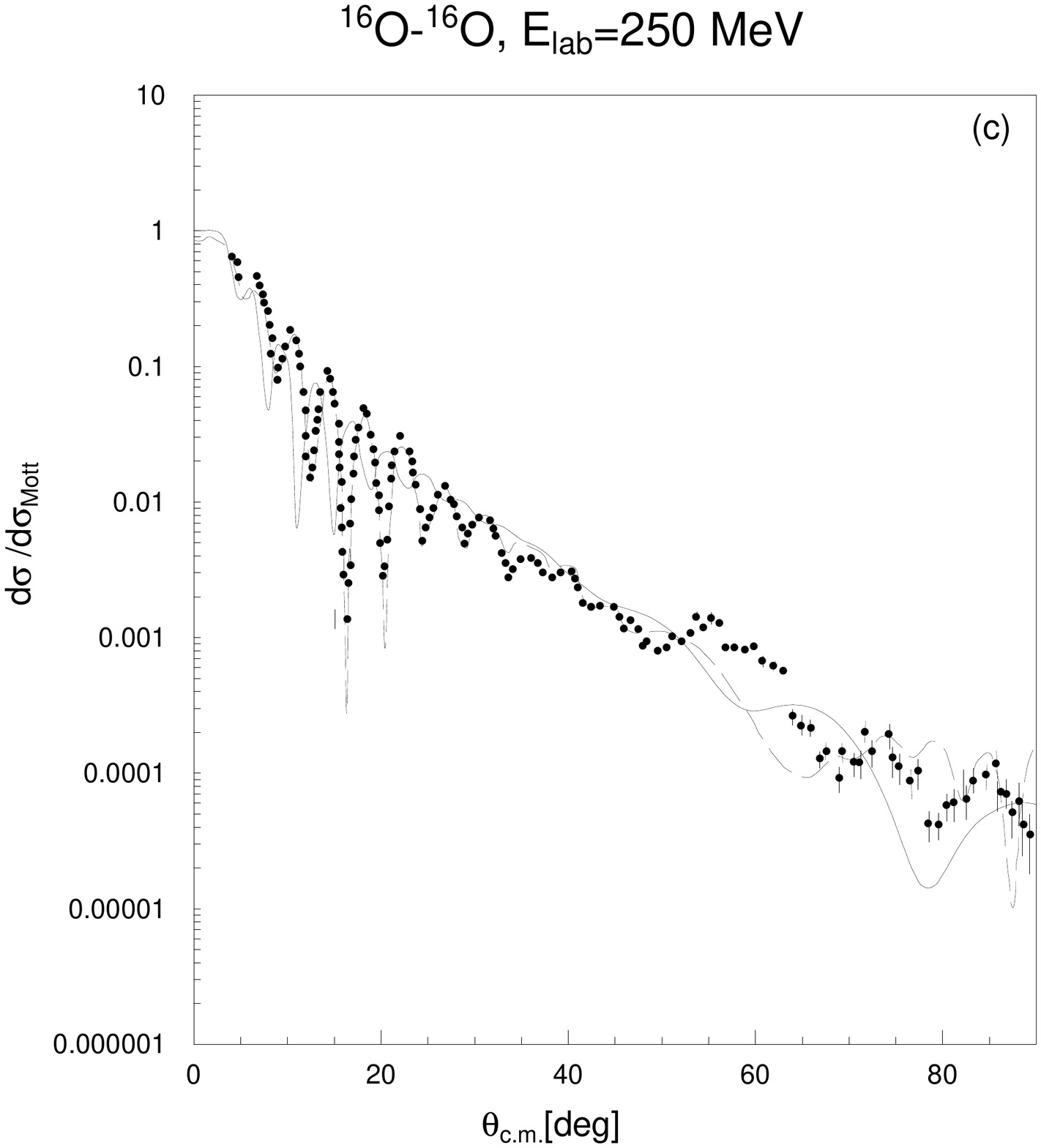}
\end{figure}
\begin{figure}[h!]
\includegraphics [scale=0.6]{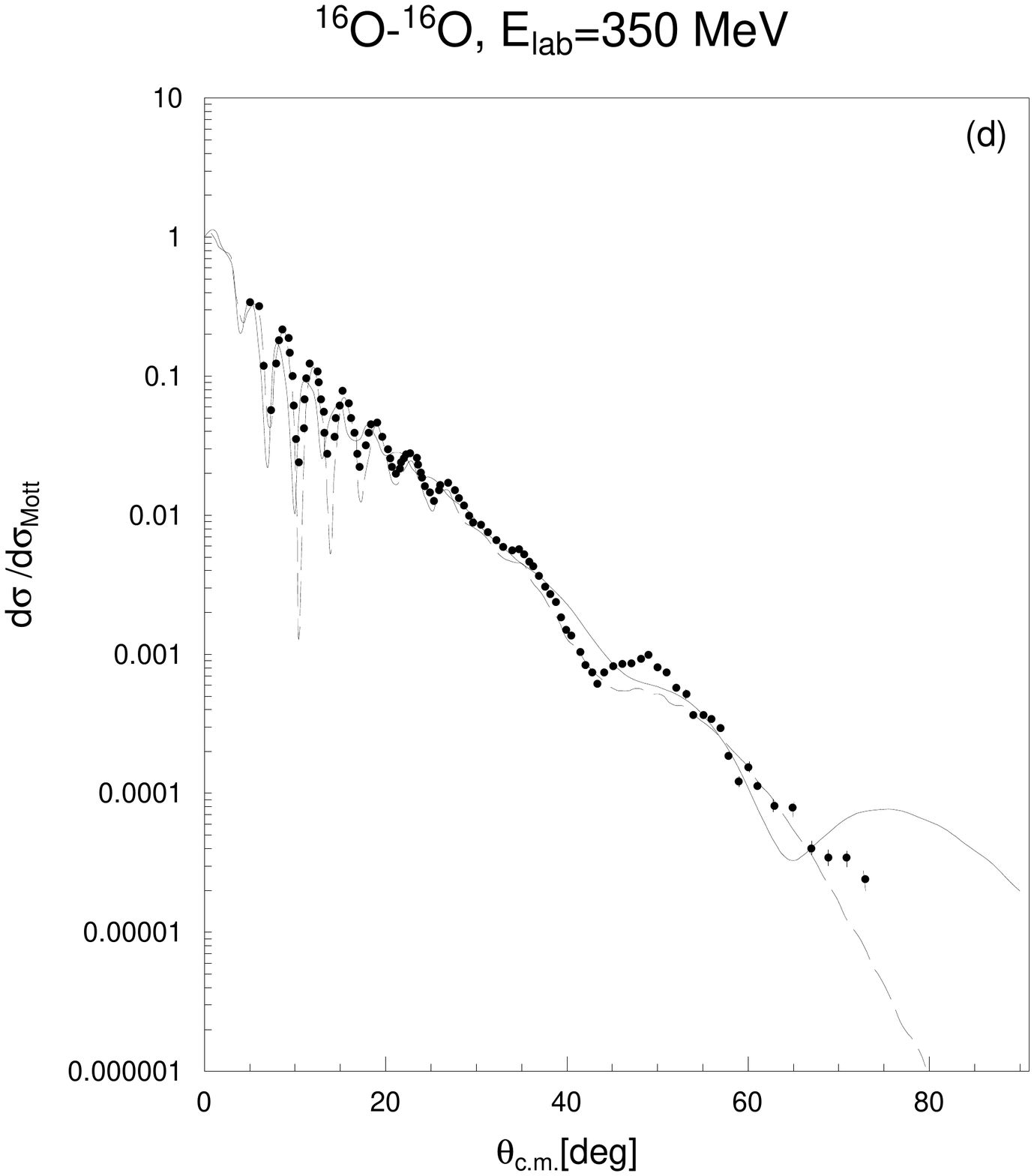}
\end{figure}
\begin{figure}[h!]
\includegraphics [scale=0.6]{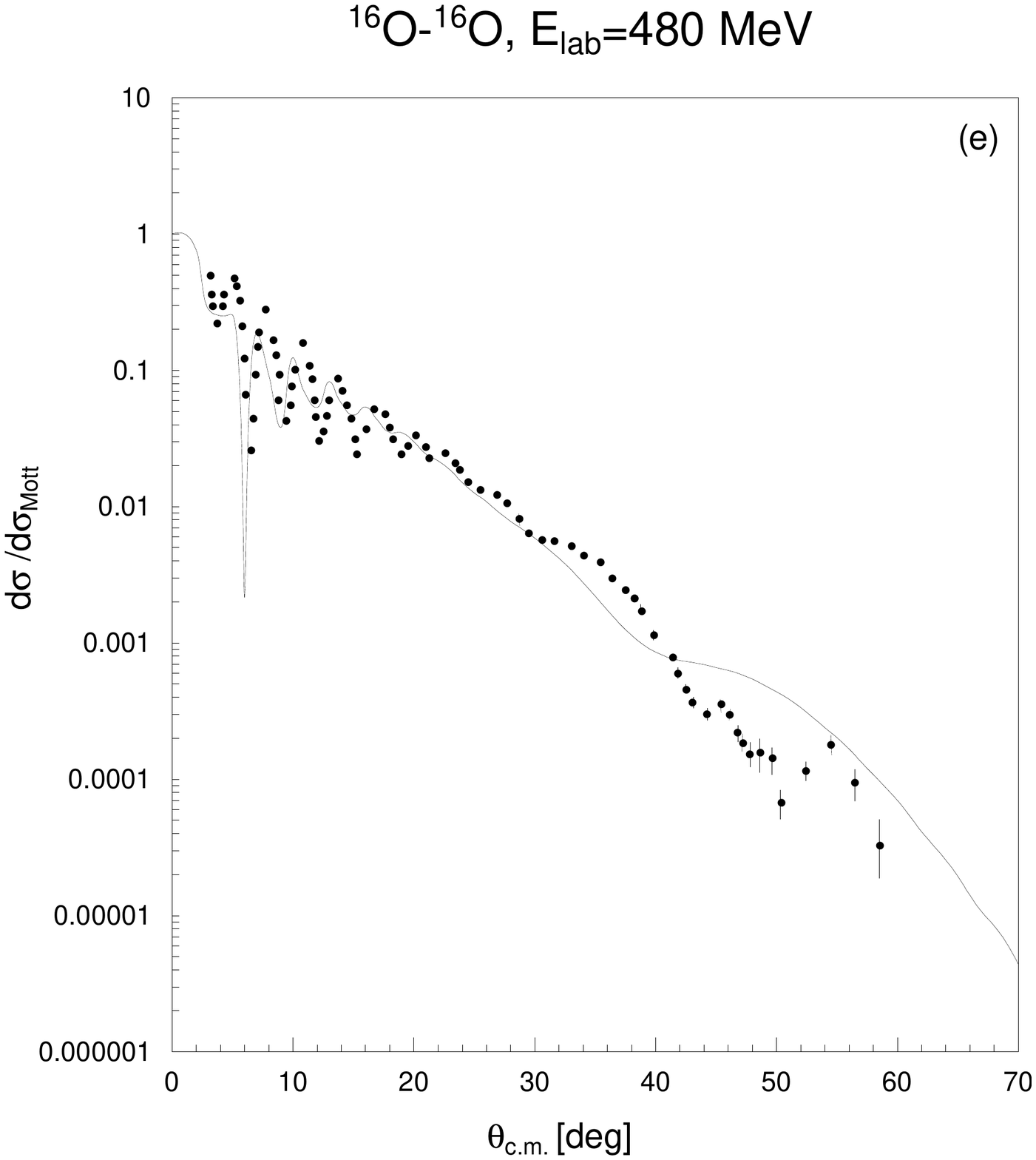}
\caption{As the figure 5  but for the $^{16}O-^{16}O $ scattering at
$ E_{lab}=124, 145, 250, 350, 480$ $ MeV $. The experimental
scattering data
 are taken from the references \cite{41,42,43,44,45}.}
\end{figure}
\newpage
\appendix*
\section{A brief introduction to the $LOCV$ formalism with the \textit{Reid}68 interaction}
In the LOCV method, we use an ideal Fermi gas type wave function for
the single particle states and the  variational techniques, to find
the wave function of interacting system
\cite{46,461,462,463,464,465}, i.e.,
\begin{equation}
\psi={\cal F}\Phi
\end{equation}
where (${\cal S}$ is a symmetrizing operator)
\begin{equation}
{\cal F} ={\cal S} \prod_{i>j} F(ij).
\end{equation}
The  correlation functions $F(ij)$ are operators and they are
written as :
\begin{equation}
F({\it ij}) = \sum_{\alpha , {\it k}} f^{\it (k)}_{\alpha}({\it ij})
O^{\it (k)} _{\alpha}({\it ij}).
\end{equation}
In above equation $\alpha = \{S,L,J,T\} $ , ${\it k}=1,3 $ and
\begin{equation}
O_{\alpha}^{\it k =1,4} =1,({2\over 3}+{1\over 6}S_{12}^{I}),
({1\over 3}-{1\over 6}S_{12}^I).
\end{equation}
In the case of the \textit{Reid}68 potential, the spin-singlet
channels with the orbital angular momentum $L \neq 0 $ and the
spin-triplet channels with $L\neq J\pm 1, {\it k} $ is superfluous
and set only to unity, while for $ L= J \pm 1 $ it takes the values
of 2 and 3. All of the channel correlation functions
$f_{\alpha}^{(1)},f_{\alpha}^{(2)}$ and $f_{\alpha}^{(3)}$ heal to
  the modified Pauli
function $f_{P}(r)$,
\begin{equation}
f_{P}(r)=[1-l(k_Fr)^2]^{-{1\over 2}}
\end{equation}
with
\begin{equation}
l(x)={3\over 2x}{\cal J}_1(x)
\end{equation}
where ${\cal J}_J (x)$ are the familiar spherical Bessel functions
and the Fermi momenta $k_F$ is fixed by the nuclear matter density
i.e., $k_F=({3\pi^2\over 2}\rho)^{1\over 3}$.

The nuclear matter energy per nucleon is \cite{461,462,463,464,465},
\begin{equation}
E_{in}=T_F+E_{MB}[F].
\end{equation}
$T_F$ is simply the Fermi gas kinetic energy and it is written as
\begin{equation}
T_F={3\over 5}{\hbar^2k_F^2\over 2m}.
\end{equation}
The many-body energy term $E_{MB}[F]$ is calculated by constructing
a cluster expansion for the expectation value of our Hamiltonian,
\begin{equation}
H=\sum_i {{p_i}^2\over 2m}+\sum_{i>j}V_{ij}
\end{equation}
where $V_{ij}$ is the bare N-N interaction. Then, we keep only the
first two terms in a cluster expansion of the energy functional:
\begin{equation}
E[F] = {1\over A} {< \Psi \vert H \vert \Psi >\over {< \Psi \vert
\Psi>}} = T_F+E_{MB}= T_F+E_2+E_3+ \ldots
\end{equation}
The
two-body energy term  is defined as,
\begin{equation}
E_2 = (2A)^{-1}\sum_{ ij}<ij\vert {\cal V}(12)\vert ij>_a\label{a1}
\end{equation}
where
\begin{equation}
{\cal V} (12) = -{{{\hbar}^2}\over
{2m}}[F(12),[\nabla^2_{12},F(12)]]+F(12)V(12)F(12)
\end{equation}
and the two-body antisymmetrized matrix element $<ij\vert{\cal
V}\vert ij>_a$ are taken with respect to the single-particle
functions composing $\Phi$ i.e. the plane-waves. In the $LOCV$
formalism $E_{MB}$ is approximated by $E_2$ and one hopes that the
normalization constraint makes the cluster expansion to converge
very rapidly and bring the many-body effect into $E_2$ term.

By inserting a complete set of two-particle state twice in the
equation ({\ref{a1}}) and performing some algebra, we can rewrite
the two-body term as following :
\begin{equation}
E_2 = E_c^{NN} + E_{\cal T}^{NN}
\end{equation}
where (c and ${\cal T}$ stand for the central and tensor parts,
respectively)
\begin{equation}
E_{\it i}^{\it j} = {2\over \pi^4\rho}\sum_{\alpha} (2T+1)
(2J+1){1\over2}\{1-(-1)^{L+S+T}\} \int_0^\infty r^2 dr{\cal
V}_\alpha^{\it i,j}(r,\rho) a_{\alpha}^{(1)^2}(r)
\end{equation}
and (${\it i}=c$ and ${\cal T} $)
\begin{equation}
{\cal V}_\alpha^{c,NN}(r,\rho)={\hbar^2 \over m}\{
f_{\alpha}^{(1)^{\prime^2}}+{m\over \hbar^2} V^c_\alpha
f_\alpha^{(1)^2}\}
\end{equation}
$$
{\cal V}_\alpha^{{\cal T},NN}(r,\rho)=\{ {\hbar^2 \over m} \{
f_\alpha^{(2)^{\prime^2}} + {m\over \hbar^2}( V^c_\alpha+2V^{\cal
T}_\alpha -V_\alpha^{LS}) f_\alpha^{(2)^2}\}
a_{\alpha}(r)^{(2)^2}+{\hbar^2 \over m}\{
f_\alpha^{(3)^{\prime^2}}$$
\begin{equation}
+{m\over \hbar^2}( V^c_\alpha-4V^{\cal T}_\alpha -2V_\alpha^{LS})
f_\alpha^{(3)^2}\} a_{\alpha}^{(3)^2}(r)+
\{r^{-2}(f_\alpha^{(2)^2}-f_\alpha^{(3)^2}+{m\over
\hbar^2}V_\alpha^{LS}f_\alpha^{(2)}f_\alpha^{(3)}
)\}b_\alpha^2\}a_{\alpha}^{(1)^{-2}}(r)
\end{equation}
\begin{equation}
a_{\alpha}^{(1)^2}(r,\rho)=I_{J}(r,\rho)
\end{equation}
\begin{equation}
a_{\alpha}^{(2)^2}(r,\rho)=(2J+1)^{-1}[(J+1)I_{J-1} (r,\rho)
+JI_{J+1}(r,\rho)]
\end{equation}
\begin{equation}
a_{\alpha}^{(3)^2}(r,\rho)=(2J+1)^{-1}[JI_{J-1}
(r,\rho)+(J+1)I_{J+1}(r,\rho)]
\end{equation}
\begin{equation}
b_{\alpha}^2(r,\rho)=2J(J+1)(2J+1)^{-1}[I_{J-1}
(r,\rho)-I_{J+1}(r,\rho)]
\end{equation}
\begin{equation}
I_{J}(r,\rho)=(2\pi^6 \rho^2)^{-1}\int_{|{\bf k_1}|,|{\bf k_2}|\leq
k_F} d{\bf k_1} d{\bf k_2}{\cal J}_J^2 (\vert {\bf k_1} -{\bf
k_2}\vert r).
\end{equation}
The potential functions $V^c_\alpha,V^{\cal T}_\alpha$,.....etc.,
are given  in the references \cite{18,19}. The calculation of $E_3$
is discussed in the reference \cite{11p} and the references therein.

The  normalization constraint as well as the coupled and uncoupled
differential equations for the NN-channels, coming from the
Euler-Lagrange equations, are similar to those were described in the
references \cite{461,462,463,464,465}.

The following important points consider in the $LOCV$ formalism: (i)
Beside the inter-particle potentials, no free parameter is used in
the $LOCV$ method, i.e. it is fully self-consistent. (ii) To keep
the higher cluster terms as small as possible, it considers the
constraint in the form of a normalization condition
\cite{461,462,463,464,465} . This was tested by calculating the
three-body cluster terms with both the state-averaged and the
state-dependent correlation functions \cite{11p}. (iii) In order to
perform an exact functional minimization of the two-body cluster
energy with respect to the short-range behavior of correlation
functions, it assumes a particular form for the long-range part of
correlation functions. (iv) Rather than simply parameterizing the
short-range behavior of the correlation functions, it performs an
exact functional minimization \cite{16p}. So, in this respect it
also saves an enormous amount of the computational time. For
example, a nuclear matter $LOCV$ calculation with the $Nijmegen$
group potentials at the given density takes a few minutes CPU time
on a 1.8 $GHz$ personal computer.

Recently \cite{15p}, it was shown that the neutron (nuclear) matter
$LOCV$ calculations with the various two-body interactions, e.g. the
$Bethe$ homework potential   and the $Argonne$ $Av^\prime_8$
interaction \cite{16p}, reasonably agree with those of $FHNC$ and
Auxiliary Field Diffusion Monte Carlo ($AFDMC$)
\cite{17p,18p,19p,20p,21p,22p} methods. Moreover, it was realized
that the different
 many-body methods such as the $LOCV$ and the fermions hypernetted chain $FHNC$ approaches give results close to each
 other when the normalization constraint is imposed in its correct form. Therefore, the
 normalization constraint plays an important role in the minimizing of the many-body terms.

So  in the $LOCV$ framework by using e.g. the Reid68 interaction, we
solve the set of
 Euler-Lagrange differential  equations to find the correlation functions. Then we can find the $SNM$-$EOS$
  by calculating the expectation value of the Hamiltonian. The minimization of the
  $LOCV-EOS$
  gives some values for the binding and saturation density of the $SNM$, which are demonstrated
  in the tables 1 and A.1.  Obviously, as it is well known one should not expect to get the exact $SNM$ empirical values.
 But in the M3Y type interactions, the situation is different, in order to ensure the empirical
  saturation density and the binding energy as well as incompressibility of the symmetric
   nuclear matter, an external density dependent factor is multiplied to the original radial M3Y
   interactions and the constants of this density dependent function are obtained such that one
   could reproduce these empirical saturation properties for the $SNM$. So the case of the $LOCV$ method
   is different from the M3Y type interactions. The separation of radial and density dependent parts of
   the $LOCV-AEI$ is done only to make it possible to use the $LOCV-AEI$ in the double folding procedure.

In the table A.1 we compare the $LOCV$ results on the saturation
properties of $SNM$ by using different
 interactions with other many body techniques (The $BB$, $BHF$, $CBF$ and $BHF-ESC$ stand for the $Brueckner$,  $Bethe$, $Brueckner$, $Hartree$, $Fock$,
 correlated-basis-function and $BHF$ using extended-soft-core interactions, see the references \cite{464}  and \cite{131},
 and the references therein,  for detail, respectively). So the
 $EOS$
 of SNM is directly calculated by the $LOCV$ formalism and there is
 no other constraint  for obtaining the saturation properties of $SNM$.

Finally we should mention that the effect of $TBF$ have been fully
discussed especially in the references \cite{461,463,464}.\\ \\

TABLE A.1: The saturation energy and the density of nuclear matter
as well as its incompressibility for different potentials and
many-body methods. See reference \cite{464} for detail.
\\
\begin{center}
\begin{tabular}{c c c c c c}
\hline \hline

 Potential & Method & Author & $ \rho_{0} (fm^{-3})$ & $ E(\rho_{0}) (MeV)$ &  $ \mathcal{K} (MeV)$  \\

 \hline

  $ AV_{18} $  &  LOCV &  BM \cite{464}&  0.310 &  -18.46 &  302   \\

  $ AV_{14} $  &  LOCV &  BM \cite{464}& 0.290  &  -15.99 &   248  \\

    & FHNC  &  WFF  \cite{464} &  0.319 &  -15.60 &   205  \\

    &  BB &  DW  \cite{464} &  0.280 &  -17.80 &   247  \\

    &  BHF &  BBB  \cite{464}&  0.256 &  -18.26 &  -   \\

 $ UV_{14} $   &  LOCV &  BM \cite{464}&  0.366 &  -21.20 &  311   \\

    &  FHNC &  CP \cite{464}&  0.349 &  -20.00 &   -  \\

    &  FHNC &  WFF  \cite{464} &  0.326 &  -17.10 &  243   \\

 $ UV_{14}+TBF $   &  LOCV & BM \cite{464} &   0.170 & -17.33 &  276   \\

    & FHNC  &  WFF  \cite{464} &  0.157 &  -16.60 &   261  \\

    &  CBF &  FFP  \cite{464} &  0.163 &  -18.30 &   269  \\

  $\triangle$-$\textit{Reid} $   & LOCV & MI  \cite{464}  &  0.258 &  -16.28 &   300  \\

 $ \textit{Reid} $   &  LOCV &  OBI \cite{464}   &  0.294 &  -22.83 &   340  \\

    &  LOCV &  MO  \cite{464} &  0.230 &  -14.58 &   238  \\

ESC    &  BHF & FSY \cite{131} & $\sim$ 0.14 & $\sim$ -12.00 &  $\sim$ 84  \\

ESC-TBA    &  BHF & FSY \cite{131} & $\sim$ 0.16& $\sim$ -14.00 & $\sim$  173  \\

ESC-TBA-Strong    &  BHF & FSY \cite{131} & $\sim$ 0.19 & $\sim$ -16.00 & $\sim$  260 \\

 Empirical   &   &   &  0.170 &  -15.86 &  (200-300)   \\

\hline \hline
\end{tabular}
\end{center}
\end{document}